\documentclass[iop,apj,tighten]{emulateapj}
\usepackage[para,online,flushleft]{threeparttable}
\usepackage{etoolbox}
\usepackage{tablefootnote}

\usepackage{tabularx}
\newcommand\setrow[1]{\gdef\rowmac{#1}#1\ignorespaces}
\newcommand\clearrow{\global\let\rowmac\relax}
\clearrow

\appto\TPTnoteSettings{\footnotesize}

\usepackage[breaklinks,colorlinks,citecolor=cyan,linkcolor=cyan]{hyperref} 

\usepackage[all]{hypcap} 
\usepackage{graphicx}
\usepackage{relsize}

\shorttitle{The Central $M/L$ of Low-Mass ETGs}
\shortauthors{Pechetti et al.}

\begin{document}
\title{Detection of Enhanced Central Mass-to-Light Ratios in Low-Mass Early-Type Galaxies: Evidence for Black Holes?}
\author{Renuka Pechetti\altaffilmark{1}, Anil Seth\altaffilmark{1}, Michele Cappellari\altaffilmark{2}, Richard McDermid\altaffilmark{3}, Mark den Brok\altaffilmark{4}, Steffen Mieske\altaffilmark{5}, Jay Strader\altaffilmark{6}}
\affil{\begin{flushleft}
\textit{{\scriptsize$^1$Department of Physics and Astronomy, University of Utah, 115 South 1400 East, Salt Lake City, UT 84112, USA\\
$^2$Sub-department of Astrophysics, Department of Physics, University of Oxford, Denys Wilkinson Building, Keble Road, Oxford OX1 3RH, UK\\
$^3$Department of Physics and Astronomy, Macquarie University, Sydney, NSW 2109, Australia\\
$^4$ETH Zurich, Switzerland\\
$^5$European Southern Observatory, Alonso de Cordova 3107, Vitacura, Santiago, Chile\\
$^6$Center for Data Intensive and Time Domain Astronomy, Department of Physics and Astronomy, Michigan State University, 567 Wilson Road, East Lansing, MI 48824, USA
} }\end{flushleft}}

\begin{abstract}
  We present dynamical measurements of the central mass-to-light ratio ($M/L$) of a sample of 27 low-mass early-type ATLAS$^{3D}$ galaxies. We consider all ATLAS$^{3D}$ galaxies with 9.7~$<$~log(M$_\star/$M$_\odot)$~$<$~10.5 in our analysis, selecting out galaxies with available high-resolution {\em Hubble Space Telescope} (HST) data, and eliminating galaxies with significant central color gradients or obvious dust features. We use the HST images to derive mass models for these galaxies and combine these with the central velocity dispersion values from ATLAS$^{3D}$ data to obtain a central dynamical $M/L$ estimate. These central dynamical $M/L$s are higher than dynamical $M/L$s derived at larger radii and stellar population estimates of the galaxy centers in $\sim$80\% of galaxies, with a median enhancement of $\sim$14\% and a statistical significance of 3.3$\sigma$. We show that the enhancement in the central $M/L$ is best described either by the presence of black holes in these galaxies or by radial IMF variations. Assuming a black hole model, we derive black hole masses for the sample of galaxies. In two galaxies, NGC~4458 and NGC~4660, the data suggests significantly over-massive BHs, while in most others only upper limits are obtained. We also show that the level of $M/L$ enhancements we see in these early-type galaxy nuclei are consistent with the larger enhancements seen in ultracompact dwarf galaxies (UCDs), supporting the scenario where massive UCDs are created by stripping galaxies of these masses.
\end{abstract}

\maketitle

\section{Introduction}
Early-type galaxies(ETGs) include all elliptical (E) and lenticular (S0) galaxies that are characterized by old, red stellar populations. Massive ETGs contain very little gas and dust, have quenched their star formations, and typically host massive black holes~(BHs) at their centers.  The BH masses in these galaxies follow relatively tight scaling relationships with overall galaxy properties, suggesting that the BHs and galaxies have co-evolved \citep[e.g.][]{kormendyandho13}. 
\par In lower mass ETGs (3$\times$10$^{9}$$\lesssim$$M_\star/M_\sun$$\lesssim$4$\times$10$^{10}$) there is also evidence that they host BHs at their centers: X-ray measurements show weak AGN at the centers of many of these low-mass ETGs \citep{gallo10,miller12}. \citet{gallomiller15} use these data to derive estimates for the BH occupation fraction, and find that most galaxies with mass above $\sim$3$\times$10$^9$~M$_\odot$ do host massive BHs. However, very few BH mass measurements exist for these galaxies.  The recent compilation of BH masses by \citet{saglia16} contains just 6 galaxies with velocity dispersions within the effective radius $\sigma_e \lesssim 100$~km~s$^{-1}$, 2 ETGs with galaxy bulge masses below 3$\times$10$^{10}$~M$_\odot$, and 10 ETGs with galaxy dynamical masses below $\sim$3$\times$10$^{10}$~M$_\odot$ \citep{atlas3d15}.  The lowest mass ETGs with dynamical BH mass estimates have $M_\star\sim$$10^9$~M$_\odot$ \citep{vandenbosch10,ngyuen17,verolme02}.

\par Using existing dynamical BH mass estimates, an empirical correlation is seen between the velocity dispersions and BH masses; the $M-\sigma$ relationship \citep[e.g.,][]{magorrian98,gebhardt2000,mcconnell_ma13,kormendyandho13,saglia16}. Although this relationship is very well constrained at the high mass end, fewer measurements exist at lower masses due to the difficulty of measuring the small BHs expected in these galaxies. \citet{lasker16} and \citet{greene16} show that late-type galaxies with $\sigma$$\sim$100 km/s (similar to the Milky Way) and bulge masses of $\sim$10$^{10}$~M$_\odot$ show a large scatter in BH mass as measured from masers. This wide range of BH masses can also be interpreted as a steepening of the scaling relationships \citep{graham15,savorgnan16}.  No comparable sample of low-mass ETG BH mass measurements is available.

\par
\begin{figure*}[ht]

\epsscale{1}
\includegraphics[width=\linewidth]{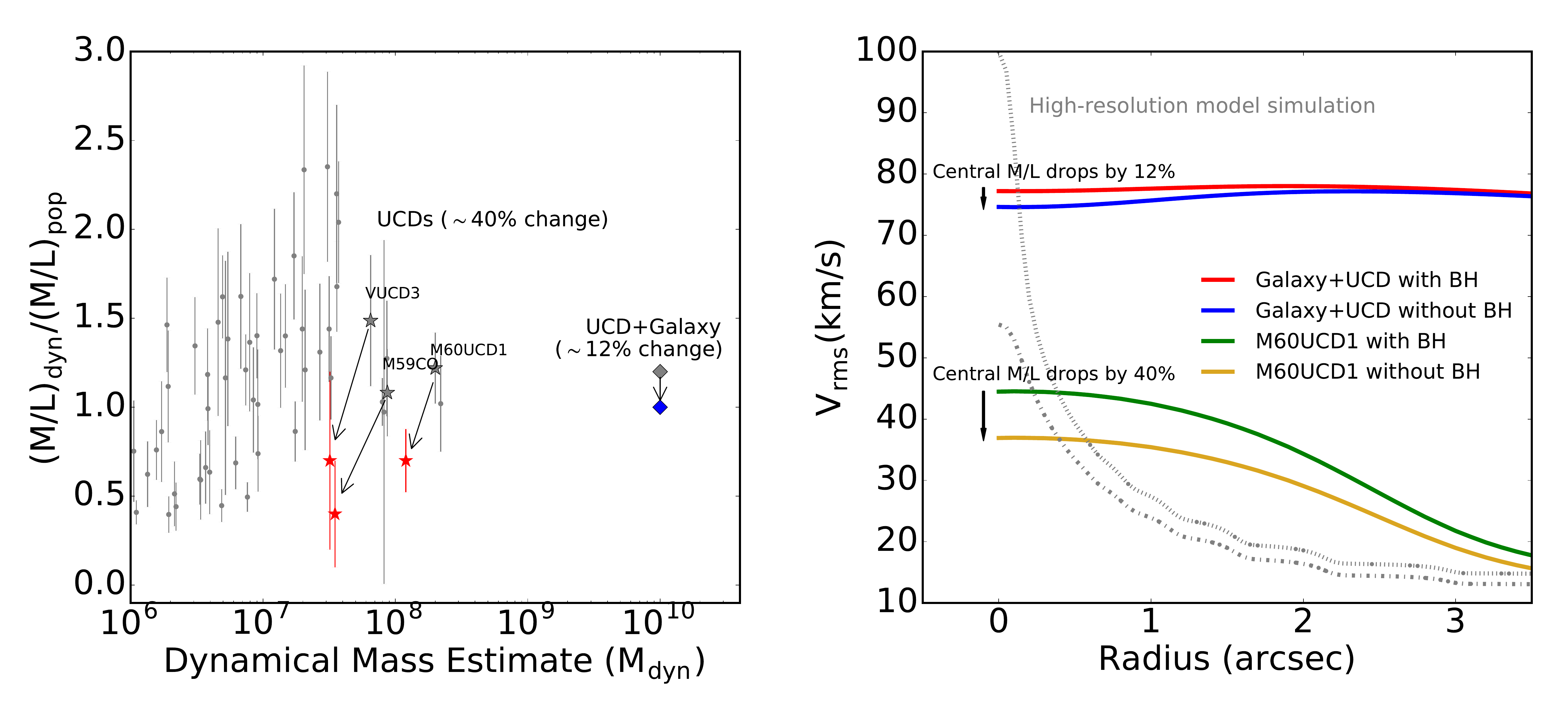}
\caption {\textit{Left Panel}: Gray circles are globular clusters and UCDs from \citet{mieske13}. Elevation in the dynamical $M/L$ \{$(M/L)_{\rm dyn}$\} is observed when compared to the stellar $M/L$ \{$(M/L)_{\rm pop}$\}. The stellar dynamical $M/L$ drops for UCDs (in red) and UCD+Galaxy model (in blue) after including the dynamical effect of a BH. \textit{Right Panel}: Simulated radial velocity dispersion profiles for M60-UCD1 with a PSF FWHM of $2''$. The bottom profiles are for M60-UCD1 only, with (green) and without (yellow) a black hole in it. The top profiles are when a galaxy is added around M60-UCD1 with (red) and without (blue) a black hole in it. The gray lines show models with adaptive optics resolution for M60-UCD1 with (top gray profile) and without (bottom gray profile) a black hole. Note the drop in the central dispersion between these high resolution models and those simulated at ground-based resolution.}
\label{fig:1}
\end{figure*}

 One promising new avenue for finding BHs in low-mass galaxies is the recent discovery of BHs in Ultracompact Dwarf Galaxies (UCDs). UCDs are dense stellar systems with masses ranging from a few million to a hundred million solar masses and sizes $< 100$~pc \citep[e.g.][]{norris14}. UCDs, like ETGs, are typically found in denser environments. Integrated dispersion measurements of the UCDs revealed dynamical mass-to-light ratio ($M/L$) estimates higher than expected based on their stellar populations \citep[e.g.][see also Figure~\ref{fig:1}]{hasegan05,mieske13}. Recently, using resolved kinematics of the UCDs, these enhanced $M/L$s have been shown in a few cases to be due to the presence of supermassive BHs \citep{seth14,ahn17}. These BHs make up a high fraction of the mass of the UCDs (10-20\%), inflating the integrated dispersion measurements. These high mass fraction BHs suggest that these UCDs are the tidally stripped remnants of once larger galaxies (likely ETGs). They also suggest it might be possible to detect BHs in ETGs through enhancements in their central $M/L$s, even at ground-based resolution.

\par Here, we analyze a sample of low-mass ETGs and derive accurate central $M/L$s using high-resolution mass models. We compare these $M/L$s to the dynamical $M/L$s determined at larger radii, and the central stellar $M/L$s inferred from their stellar populations. In section 2, we discuss the connection between ETGs and UCDs to set expectations for our study. Then, in Section 3 we discuss the selection of our sample of 27 ETGs from the ATLAS$^{3D}$ survey. Section 4 describes the modeling of HST data including the derivation of Point Spread Functions (PSFs), creating multi-Gaussian expansion mass models for HST images and analyzing their color maps. Section 5 describes the derivation of central $M/L$ for the galaxies using Jeans anisotropic modeling and various sources of errors associated with our models. In Section 6, we describe our results, focusing on the enhancement in $M/L$. We consider possible interpretations of this enhancement in Section 7, estimating BH masses, discussing the possibility of radially varying IMF and the impact of anisotropy in the galaxies. We conclude in Section 8.
\section{Early-Type Galaxies in the Context of UCDs}
\label{sec:ucds}
\par To motivate the measurement of central $M/L$s in ETGs, in this section we consider what the expected dynamical signature of the black holes found in UCDs would be in a typical ETG before they are stripped. As mentioned in the introduction, the dynamical $M/L$s of UCDs are higher than expected from their stellar populations. More specifically, integrated dispersion measurements of UCDs have been used to calculate dynamical $M/L$s \{$(M/L)_{\rm dyn}$\} under the assumption that the mass distribution traces the light distribution of the UCD.
These are enhanced relative to the maximum estimate of their stellar population $M/L$ \{$(M/L)_{\rm pop}$\} assuming a Chabrier initial mass function (IMF) \citep{hasegan05,dabringhausen09,dabringhausen10,frank11,mieske13,strader13}. The left panel of Figure~\ref{fig:1} shows that the enhancement can be seen in a majority of UCDs with dynamical mass estimates above a few million solar masses which show $(M/L)_{\rm dyn}$/$(M/L)_{\rm pop} > 1$. Taking into account the measured BH masses in the three objects with adaptive optics measurements, we find that $(M/L)_{\rm dyn}$/$(M/L)_{\rm pop}$ of the stellar component decreases on average 40\% relative to the integrated estimates without a BH in the recent discoveries like M60-UCD1 (1.222~$\rightarrow$~0.7) \citep{seth14}, VUCD3 (1.02~$\rightarrow$~0.7) \& M59CO (1.04~$\rightarrow$~0.4) \citep{ahn17}.

\par To examine the effect of the BH on the dispersion profiles of the UCDs, we take the measured stellar $M/L$ of M60-UCD1 and its stellar mass model \citep{seth14} to predict the dispersion profiles with and without a $2.5 \times 10^7$~M$_\odot$ BH (right panel, Figure~\ref{fig:1}) using isotropic Jeans modeling, described further in Section~\ref{sec:jam}. The original M60-UCD1 measurement was made with Adaptive Optics and thus has a resolution of $\sim$0.2$''$; but we use ground-based measurements ($2''$ full width at half maximum (FWHM) and a pixel size of $0.94''$) for our models, similar to those presented in this paper. As expected, the inclusion of a BH results in a significant difference of 17\% in the central dispersion, even at this lower resolution. Note that in Figure~\ref{fig:1}, the low-resolution model central velocity dispersions are lower (green \& yellow profiles) compared to M60-UCD1's observed dispersion \citep{strader13}. We also show models with and without a black hole observed at adaptive optics resolution (gray profiles); the upper of these profiles closely matches the observed $V_{rms}$ profile of M60-UCD1 by \citet{seth14}.

\par Based on the BH mass and nuclear properties, \citet{seth14} suggests that M60-UCD1 had a progenitor galaxy of mass around $10^{10}$~M$_{\odot}$ before it was stripped. To simulate what this progenitor would have looked like, we take the central S\'ersic component \citep[likely the former nuclear star cluster of the galaxy][]{pfeffer13}, and add a typical $10^{10}$~M$_{\odot}$ ETG around it with sersic index, $n$=2 \& effective radius, $R_{eff}$~=~1.15~kpc \citep[e.g.][]{kormendy09}. We create a mass model from these two components, and again use Jeans isotropic models to simulate the galaxy with and without a $2.5 \times 10^7$~M$_\odot$ BH. A smaller difference of 4.4\% is seen in the central dispersion than in the ``naked'' UCD simulation, due to the mixing of galaxy light from larger radii. Thus the dynamical effect of a BH is more visible in a stripped nucleus than it is in its original galaxy.
  When we translate the change in velocity dispersions to a change in $M/L$ using our dynamical models, we find that the UCD model has a drop of $\sim$40\% in central $M/L$ when a BH is included, whereas UCD+galaxy model has a drop of $\sim$10-12\% in central $M/L$ when a BH is included in the model. We also ran a similar simulation for M59CO \citep{ahn17} using a lower mass galaxy, with similar results.  

\par These simulations suggest that if comparable BHs to those found in UCDs are present in low-mass ETGs, they could inflate the inferred central $M/L$s by $\sim$10\% in ground-based resolution spectra, and thus may be detectable in existing data sets. This motivates the study of central $M/L$ measurements presented here.

\section{Data \& Sample Selection}
In this study, we combine central velocity dispersion values of the galaxies from the ATLAS$^{3D}$ survey \citep{atlas3d1}, with high-resolution mass models derived from HST images to obtain the central mass-to-light ($M/L$) ratios of these galaxies. Here we discuss the ATLAS$^{3D}$ data, and the sample selection of galaxies from the ATLAS$^{3D}$ survey with existing HST archival data.\\ \\

\subsection{ATLAS$^{3D}$ spectra}
\par We use the optical integral-field spectroscopic observations for the ATLAS$^{3D}$ galaxies taken from SAURON Integral Field Unit (IFU) on the William Herschel Telescope (WHT). The SAURON IFU \citep{bacon01} has a spatial sampling of $0.94''$ pixel$^{-1}$in a low-resolution mode with a spectral window of 4810 - 5350 \AA. The ATLAS$^{3D}$ team has used the resulting 3D data cubes to derive kinematic measurements of galaxies \citep{atlas3d1}\footnote{\tt Available from http://purl.org/atlas3d}.  We use two data products from ATLAS$^{3D}$:\\
1) The central velocity dispersions of the galaxies.\\
2) The fluxes of the spectra to fit the PSF of the ATLAS$^{3D}$ data.\\
For one galaxy, NGC~4342, ATLAS$^{3D}$ data was taken under poor conditions resulting in a less reliable data, so we use data from \citet{vandenbosch98} for this galaxy. Its spectra is from WHT and the resolution is $\sim1''$; similar to ATLAS$^{3D}$ data.

\subsection{Sample selection from ATLAS$^{3D}$ galaxies}
\par The ATLAS$^{3D}$ project is a multi-wavelength survey combined with numerical simulations and theoretical modeling of galaxy formation for 260 ETGs. These are selected to be within a volume of $1.16\times10^5$~Mpc$^3$ and a radius of 42~Mpc consisting of morphology of ellipticals (E-type) and S0 galaxies. Detailed selection criteria for these 260 galaxies, which are a complete and representative sample of the nearby ETG population, is described in \citet{atlas3d1}. 

\par
\begin{figure}

\epsscale{1.2}
\plotone{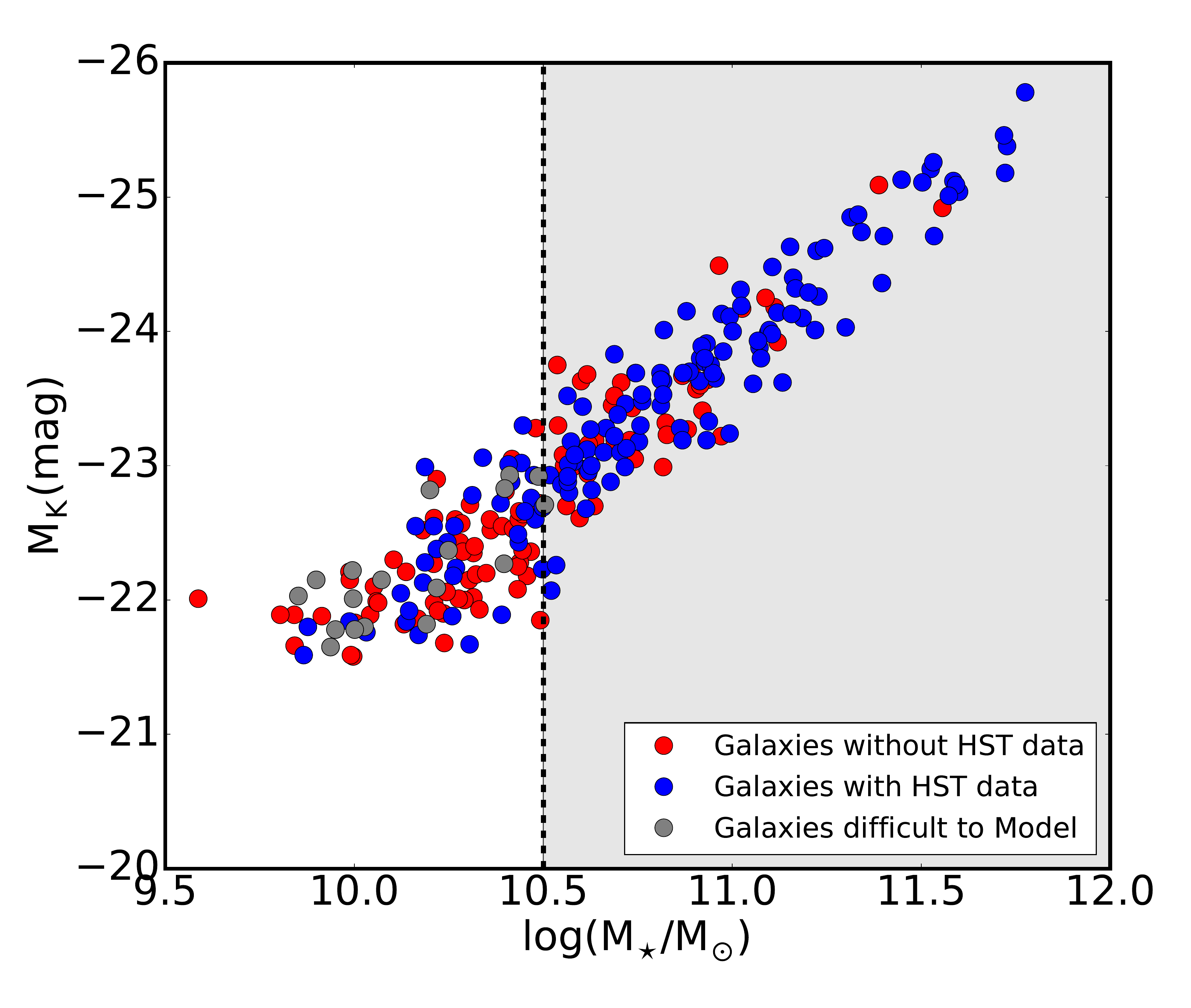}

\caption{The ATLAS$^{3D}$ galaxy sample \citep{atlas3d1}. Our selected low-mass ETG sample has log$M_{\star}<10.5$~M$_\odot$.  The $M_\star$ is the total stellar mass measurement of obtained by multiplying the total luminosity of the galaxy with the stellar $M/L$ determined dynamically within $\sim$1 effective radius ($R_e$) \citep[Table 1 of][]{atlas3d15}.  The galaxies colored in blue have HST data} and those colored in red don't have HST data. Gray points are galaxies that had visible dust lanes or saturated pixels and were difficult to model, thus were removed from our analysis.
\label{fig:2}
\end{figure}

 This paper focuses on the lowest mass portion of the sample of galaxies from ATLAS$^{3D}$ survey, as existing black hole demographic studies include very few ETGs in this mass range. The ATLAS$^{3D}$ galaxies were selected to have $M_K<-21.5$ mags (log$M_{\star}\gtrsim9.7$~M$_{\odot}$), which sets the lower mass limit of the sample.  We then make an upper mass cut at a dynamical mass of log$M_{\star}<10.5$~M$_{\odot}$ to select the low-mass galaxies; this gives us 115 galaxies of which we analyze only a subsample (see Figure~\ref{fig:2}).

\subsection{Sample selection from HST survey}
To create accurate mass models for measuring a central dynamical $M/L$, we require imaging data at higher resolution than the kinematics. We therefore selected only those galaxies that had archival HST data for the ETGs. We downloaded the HST imaging using the Hubble Legacy Archive\footnote{\tt Available from https://hla.stsci.edu} including imaging from the Wide-Field Planetary Camera 2 (WFPC2), Advanced Camera for Survey (ACS)/ Wide Field Channel (WFC), ACS/High Resolution Camera (HRC), and Wide Field Camera 3 (WFC3)/Ultra Violet and Visible light(UVIS).

\par To accurately model the light emitted from a galaxy, we used data only from filters with central wavelengths $\geqslant$~4739\AA/ F475W. This minimizes any uncertainties due to dust and stellar population variations. From the HST data, we found a total of 55 out of 115 galaxies that had usable data in the preferred wavelengths. By visual inspection of these 55 galaxies, we discarded those that had noticeable dust lanes or saturated pixels at the center, as they would be difficult to model, leaving us with 35 galaxies. We then performed an analysis of color gradients of the galaxies to identify color variations at their centers, as described in Section~\ref{sec:color}. Those with large color variations were discarded from the sample. This left us with a sample of 27 galaxies in which we can determine the accurate central $M/L$s. Figure~\ref{fig:2} shows our sample selection from ATLAS$^{3D}$ and HST data, including the galaxies that were discarded. A complete list of these 27 galaxies that were used for modeling is available in Table~\ref{table:mlr}, Appendix.

\section{Creating high-resolution mass models}
\label{sec:mges}
By focusing on galaxies with existing high-resolution HST archival data, we can create the mass models needed to obtain accurate central $M/L$s. We create these using a Multi-Gaussian expansion (MGE) model \citep{emsellem94,cappellari02}. In this section, we first discuss how we determine the HST PSFs to enable us to deconvolve our mass models. The construction of these mass models is then described. We also discuss the effects of color gradients at centers of the galaxies and how they affect our mass models.

\subsection{Determining PSF of the HST images}

PSF modeling is important for our data as we are focusing on the central arc second of the image where its effects are significant, even at HST resolution. We derived $3''\times3''$ PSFs for HST images at the location of the galaxy nuclei using the \texttt{Tiny Tim} PSF modeling tool \citep{kristhook11}, assuming a blackbody of 4000~K. For each object, an undistorted PSF was generated using the appropriate camera and filter. For the ACS/WFC camera, we convolved a charge diffusion kernel corresponding to each filter\footnote{as described in Avila, R., et al. 2016, ACS Instrument Handbook, Version 15.0}; for WFPC2/PC camera charge diffusion was not significant and so no correction was made. For WFC3/UVIS, we used a distorted PSF from \texttt{Tiny Tim} which accounts for the charge diffusion kernel. We then fit these PSFs using \texttt{mge\_fit\_sectors} code \citep{cappellari02}, which is a 2-D MGE model \citet{bendinelli91} and decomposes the image into a series of 2-D Gaussians. The MGEs of the PSFs were forced to be circular, as our dynamical modeling code can only use symmetric PSFs.   The PSF MGEs were then used in the creation of the mass models of the galaxies.

\par To test the sensitivity of our mass models to the PSF, we also remade our mass models using an empirical PSF for galaxies with ACS/WFC data. It was derived using stars in an ACS/WFC image. We found that our final derived $M/L$ values are not sensitive to the exact PSF model we use, with $<$1\% differences.  This is small relative to the $M/L$ errors from other sources, described in Section~\ref{sec:errors}.

\subsection{Surface photometry using MGE's}
To parameterize the stellar mass distribution of the galaxy at high-resolution, we used our HST data to build a mass model for the galaxy. We built the mass models using an MGE decomposition python code and fitting method described by \citet{cappellari02}. We used the code to first find the galaxy's center, its orientation in the image, and then used the information to divide the galaxy into equally spaced sectors in eccentric anomaly. Using these sectors, Gaussians consisting of various widths were fit to a set of galaxy surface brightness measurements that were calculated along those sectors. The MGE galaxy model was then fit including convolution of the PSF MGE discussed in the previous section. The best-fit MGEs can then be deprojected to obtain a 3-D luminosity model.
\begin{figure}[ht]
\epsscale{1.25}
\plotone{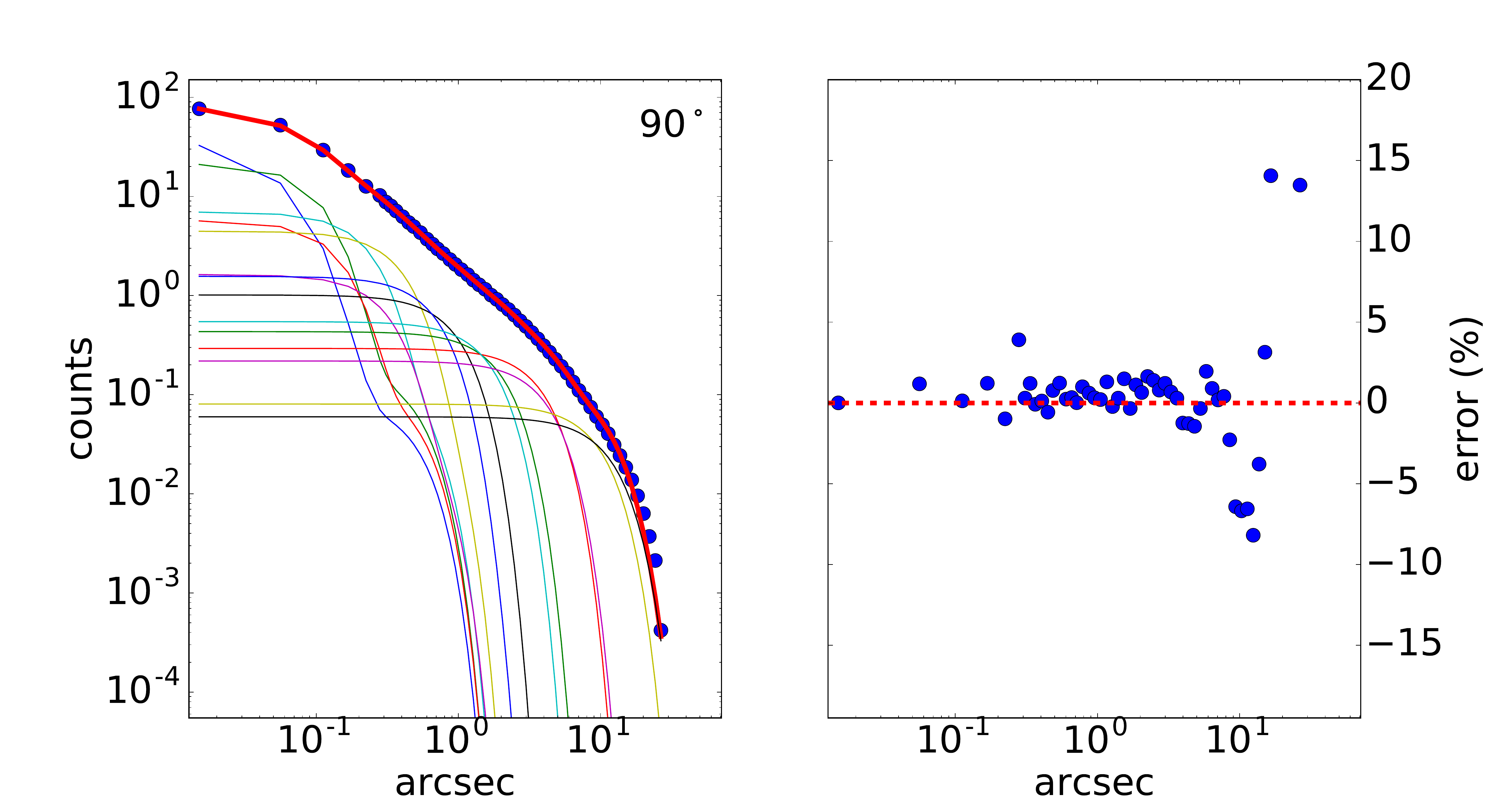}
\epsscale{1.0}
\plotone{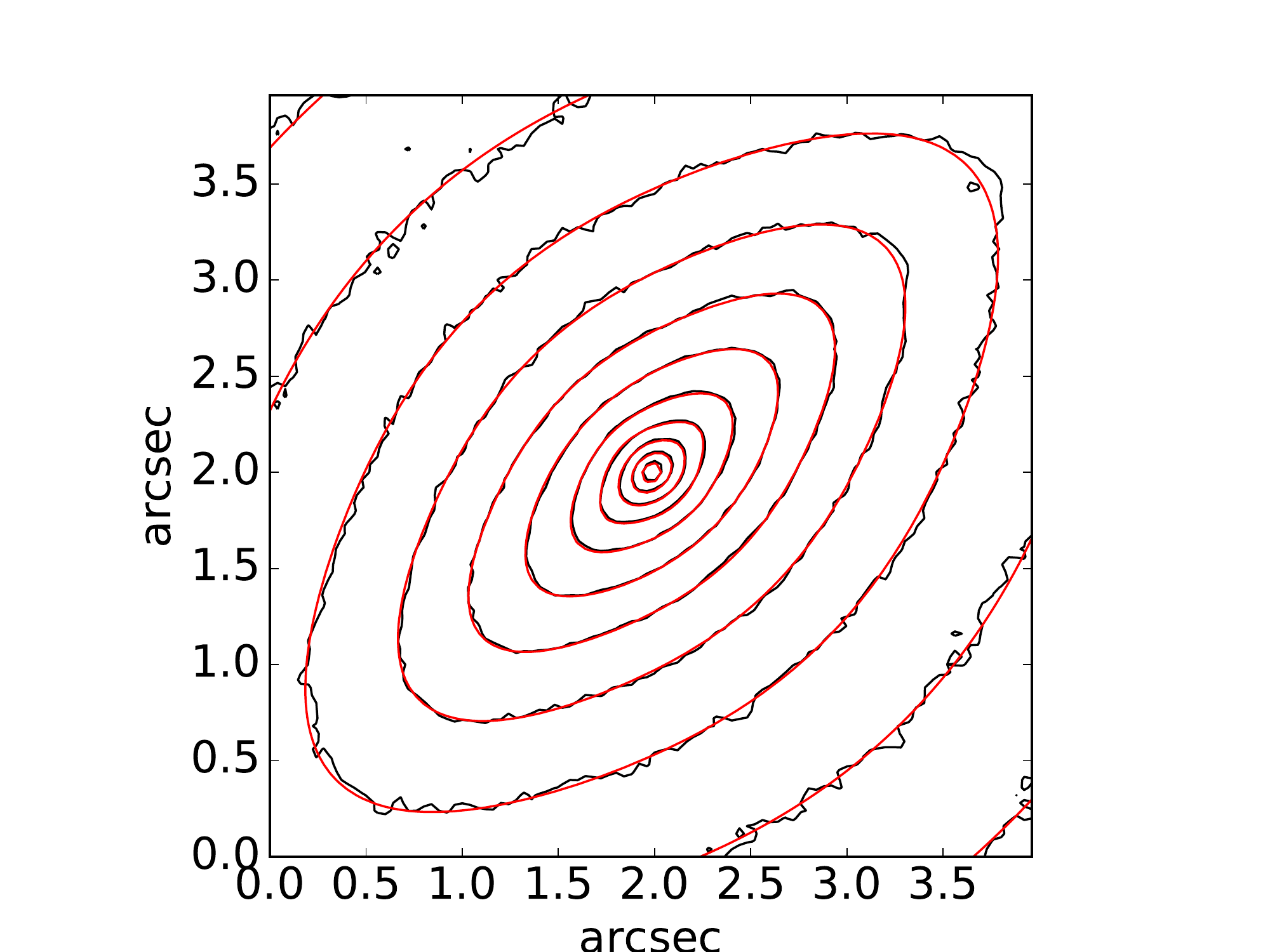}

\caption {\textit{Top panel}: Comparison between the WFC3/UVIS photometry of NGC~3522 in F814W (blue circles) and the corresponding best-fit MGE model with 14 Gaussians (solid lines) along with the residuals in the right panel.  \textit{Bottom panel}: Surface brightness contours of the central $4'' \times 4''$ of the same galaxy (black) with the MGE model overlaid in red.}
\label{fig:3}
\end{figure}

\par In nearby galaxies, sky subtraction in drizzled images often results in significant over-subtraction as the sky regions contain significant amount of galaxy light. Therefore, we add back the subtracted sky to the image of the galaxy. This leaves some low level of sky in the image, but this level is insignificant compared to the surface brightness at the centers of our galaxies. We also exclude the pixels in the image where there is no data. The MGEs were fit typically over an area of $25'' \times 25''$. A typical example of MGE fit to the surface brightness is shown in Figure~\ref{fig:3}. The derived MGE parameters of the same galaxy are given in Table~\ref{table:mge}  that describes the luminosity, width and axis ratio(q) of each Gaussian.

\begin{table}
\centering

\caption{Multi-Gaussian Expansion (MGE) parameters of all 27 galaxies}
\label{table:mge}
\def\arraystretch{1.5}
\begin{threeparttable}
\begin{tabular}{lllll}
\hline\hline

Galaxy & Filter & log$I$ & log$\sigma$ & q\\
&&($L_{\odot}$pc$^{-2}$) & (arcsec)   \\ 

\hline
... & ... & ... & ... & ...\\
NGC 3522 & F814W &5.486                          & -1.478      & 0.63 \\
&&4.709                                 & -0.861     & 0.55 \\
&&4.184                                   & -0.518     & 0.56 \\
&&3.558                                  & -0.211     & 0.72\\
&&3.804                                 &  0.096    & 0.32 \\
&&3.236                                & 0.132     & 0.79 \\
&&3.266                                & 0.228     & 0.37 \\
&&3.028                                & 0.539     & 0.66 \\
&&2.545                                 &  1.069    & 0.50 \\
  ... & ... & ... & ...&...\\                         
\hline
\\

\end{tabular}
\begin{tablenotes}
$Note$: The parameters are the luminosity (I), width($\sigma$) and axis ratio(q) of each Gaussian for NGC~3522. Only a portion of this table is shown here to demonstrate its form and content. A machine-readable version of the full table is available.
\end{tablenotes}
\end{threeparttable}
\end{table}

All derived MGE profiles of the galaxies are available in the online version of Table~\ref{table:mge}.
The MGEs are used in deriving $M/L$ of the galaxies via dynamical modeling. Given the lack of color gradients and dust in most of our galaxies, we assume that the stellar mass follows the stellar light; we examine this assumption in greater detail in the next subsection. 
\begin{figure}[ht]

\epsscale{1.3}
\plotone{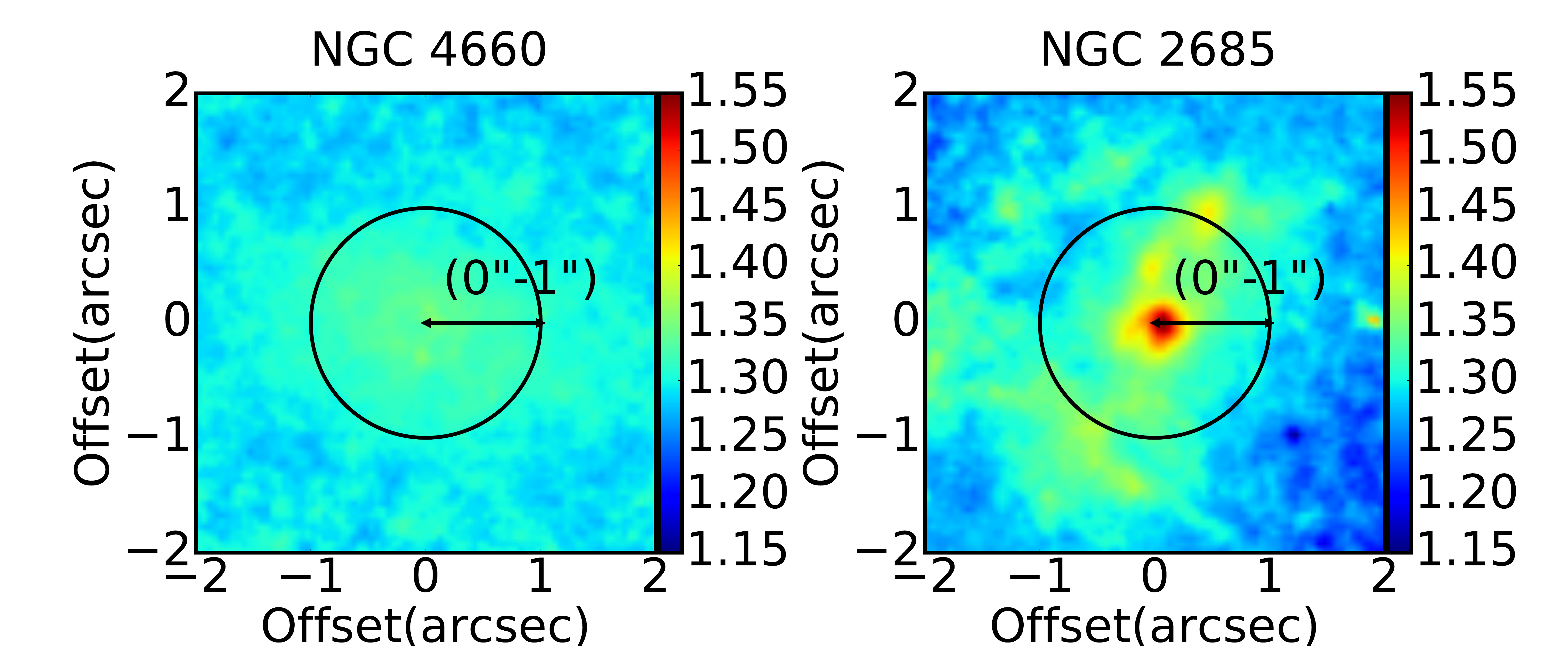}
\plotone{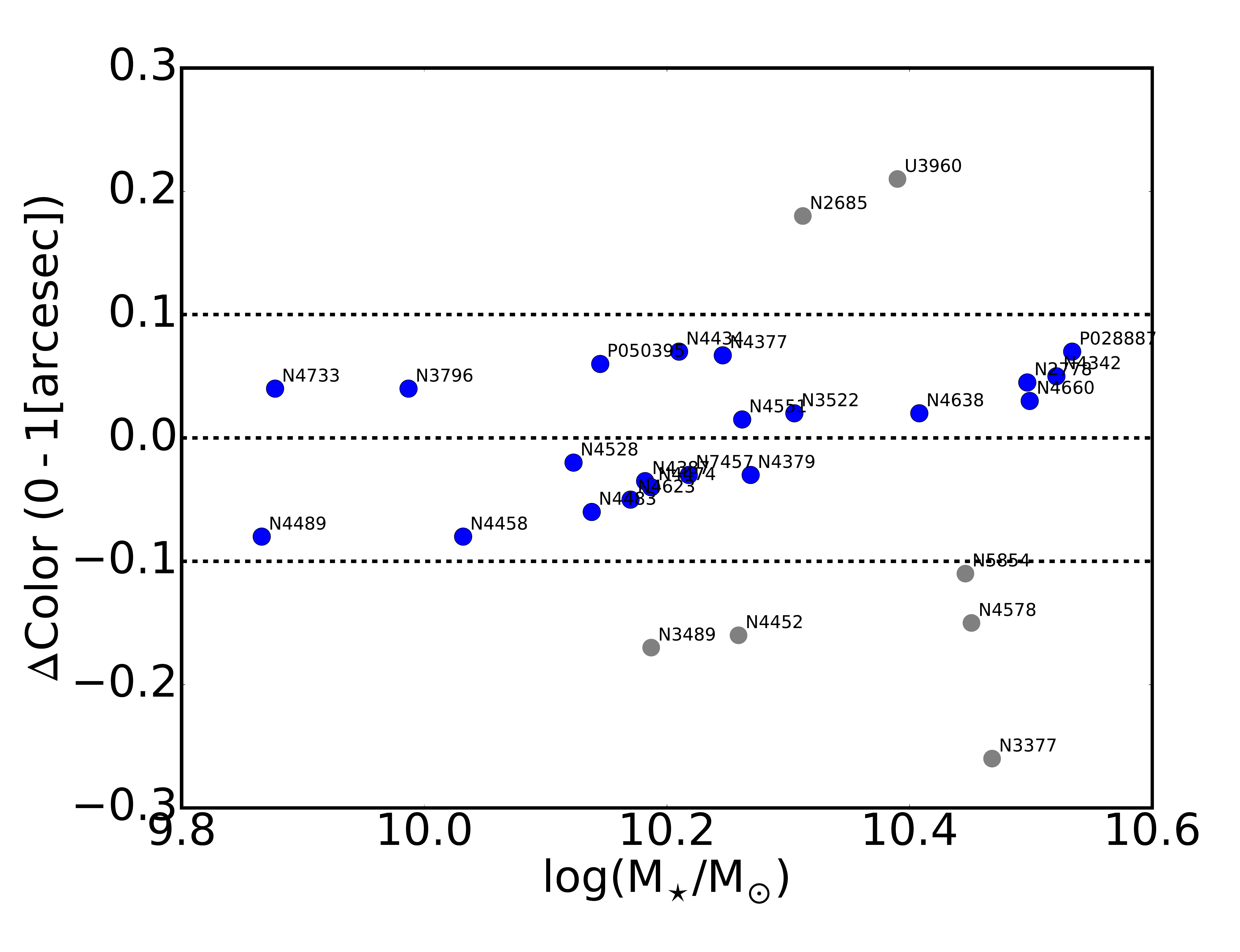}
\caption {Color gradients in our low-mass ETG sample galaxies.  \textit{Top left panel}: Color map (F555W - F814W) of central $1''$ region of WFPC2/PC image of NGC~4660, which shows a minimal color gradient within the central $1''$.  \textit{Top right panel}: Color map (F555W - F814W) of NGC~2685, which shows a significant central color gradient. \textit{Bottom panel}: Color difference in the central 1'' of the full sample of galaxies is calculated and plotted. We note that each galaxy in gray has color gradient $>$ 0.1~mags at its center; these galaxies have been excluded from most additional analysis due to the lack of reliability in the central $M/L$ estimates created by these central color gradients.}
\label{fig:4}

\end{figure}
\subsection{Determining color profiles of the galaxies}
\label{sec:color}
ETGs are often assumed to have uniformly old populations with little gas and dust \citep{thomas05}, however, especially low-mass ETGs are observed to have some stellar population gradients \citep[e.g.][]{rose05,goddard17}. Because our determination of central $M/L$s depend critically on the assumption that mass-traces-light near the center of the galaxy, we check whether there are indications of dust or color-gradients in our sample using color maps for the 27 out of 35 galaxies with data available in more than one HST filter.  

\par In this section, we quantify the color-gradients in our galaxies and examine the effects they have on our central dynamical $M/L$ estimates. Our central dispersion estimates from ATLAS$^{3D}$ are taken using $0.94''$ pixels under seeing conditions of $\sim$$2''$. Therefore, we are primarily concerned with color gradients within the central $\sim$$1''$ of the galaxy. If such color gradients exist, it would violate the assumption we are making that the $M/L$ is constant over the region we fit the kinematics.

\par To create color maps for our sample galaxies, we used F555W and F814W filters for WFPC2/PC data, F475W and F850LP filters for ACS/WFC and F475W and F814W filters for WFC3/UVIS data. To remove the PSF effects in color maps, we cross convolved the images in one filter with the PSF of the other filter before making our color maps. Example color maps for two galaxies in our sample are shown in the top panel of Figure~\ref{fig:4}. We calculated the color difference between the value measured at the center (within a circle of R = 0.05$''$), and within a circular annulus of radius R = 1$''$ and width 0.05$''$  for all the galaxies in our sample to quantify their color gradients. By visual examination, we found this difference in colors to be a good indicator of significant color variation in the galaxy nucleus.  The bottom panel of Figure~\ref{fig:4} shows this color difference for all galaxies with available color maps; for simplicity we plot the different cameras/color differences on a single plot.  

\par We exclude galaxies with large color differences in the central $1''$ because dynamical modeling would lead to inaccurate estimation of their $M/L$s. For example, if a galaxy is redder at the center than at $1''$, we will underestimate the relative amount of mass at the center, and this will lead to an overestimation of the dynamical $M/L$. To quantify this effect, we alter our MGE models to find the error in $M/L$ corresponding to a color gradient in the central $1''$. We took our MGE models, and varied the $M/L$ of each individual component using the color-$M/L$ relation of \citet{bc03}. We converted the constant mass-follows-light profile to a changing mass profile based on the color of the galaxy. We found that for color gradients of 0.1 mag (in F555W-F814W, where the color corresponds to the largest difference in stellar $M/L$), our best-fit dynamical $M/L$s vary by 3-4\%. This error is comparable to the contribution of the dispersion error to our dynamical $M/L$s, and therefore we remove all galaxies with color gradients more than 0.1~mags from the quantitative analysis presented in later sections.

For galaxies without color information, we examined their spectra for emission line contribution, as this could also result in inaccurate mass profiles.  From this examination, we removed NGC~5173 as it had significant emission line flux in its ATLAS$^{3D}$ spectra. The remainder of the sample of 27 galaxies (22 with color maps and 5 without color maps) is used for deriving the $M/L$ using dynamical models.

\subsection{Outer to Central stellar population variations}
\label{sec:inner_outer}
In assessing our central $M/L$s, we compare our dynamical $M/L$s of the galaxy nuclei to those derived while fitting the full SAURON field-of-view, which extends to a typical radius of about $1R_e$ \citep{atlas3d20}. Here we explore the variations we expect due to stellar population differences.

In low-mass ETGs, there can be populations of stars of various ages \citep[e.g.][]{atlas3d30}. These stellar population variations can lead to $M/L$ gradients within a galaxy. To quantify the size of those variations in our sample, we compare the stellar population mass-to-light ratio outside the nucleus within $1 R_e$ \{$(M/L)_{\rm pop}^{\rm out}$\} with the central stellar population mass-to-light ratio \{$(M/L)_{\rm pop}$\}. We calculated the $(M/L)_{\rm pop}^{\rm out}$ by fitting the full observed spectra using a linear combination of single stellar population synthetic spectra which were created assuming a Salpeter IMF, as explained in more detail in \citet{atlas3d30}. The central $(M/L)_{\rm pop}$ values were derived by fitting only the central spectrum of the ATLAS$^{3D}$ SAURON field in an identical way to that presented in \citet{atlas3d30}. 

\begin{figure}[ht]

\epsscale{1.25}
\plotone{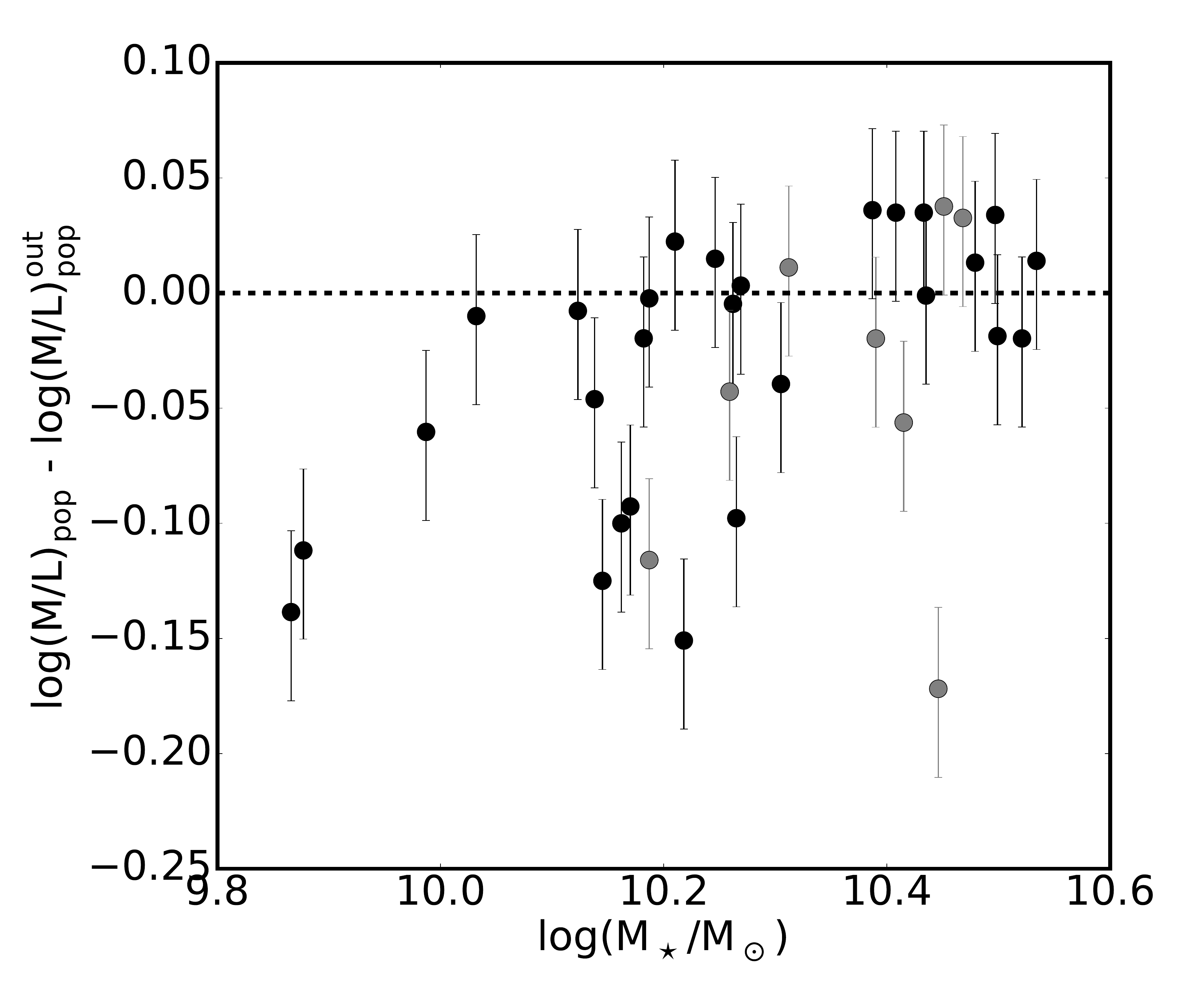}
\caption {Ratio between $(M/L)_{\rm pop}$ (central stellar $M/L$) and $(M/L)_{\rm pop}^{\rm out}$ (outer stellar $M/L$; within $1 R_e$) estimated from stellar population synthesis modeling as described in \citet{atlas3d30}. Galaxies in gray are those with identified color gradients within 1\arcsec.  Negative values indicate lower $M/Ls$ at the center.}
\label{fig:5}
\end{figure}

Figure~\ref{fig:5} shows the plot of the ratio between $(M/L)_{\rm pop}$ and $(M/L)_{\rm pop}^{\rm out}$.  We note that we calculated these values assuming both a Salpeter IMF \citep{salpeter55} and a Kroupa IMF \citep{kroupa01}; the Salpeter IMF is used for comparison to $(M/L)_{\rm pop}^{\rm out}$ for consistency, but the Kroupa $(M/L)_{\rm pop}$ values will be used later in the paper.  For most galaxies, the $M/L$ at the center of the galaxy is lower than the outer parts of the galaxy, by as much -0.15 dex.  This suggests most low-mass ETGs have younger stellar populations concentrated towards their centers. This is in agreement with the difference in the mass-weighted ages from spectral fits \citep{atlas3d30}. At the higher mass end (log$M_\star~>~10.2$), some galaxies have higher central $M/L$s, but the increase is always $<$0.04 dex. This suggests that based purely on stellar population differences, we would expect the dynamical $M/L$s at the center to typically be lower than the dynamical $M/L$ at larger radii. We will return to this result when we compare our central and outer dynamical $M/L$ values in Section~\ref{sec:compareml}.

\section{Dynamical Modeling}
\label{sec:dynamicalmodeling}
In this section, we present the estimation of the central dynamical $M/L$ of our sample galaxies. These central $M/L$s are determined using dynamical models that scale the mass models to match the measured central dispersion values from ATLAS$^{3D}$. We also discuss the corresponding errors and their detailed analysis.
\begin{figure*}[ht]

\epsscale{1.25}
\includegraphics[width=\linewidth]{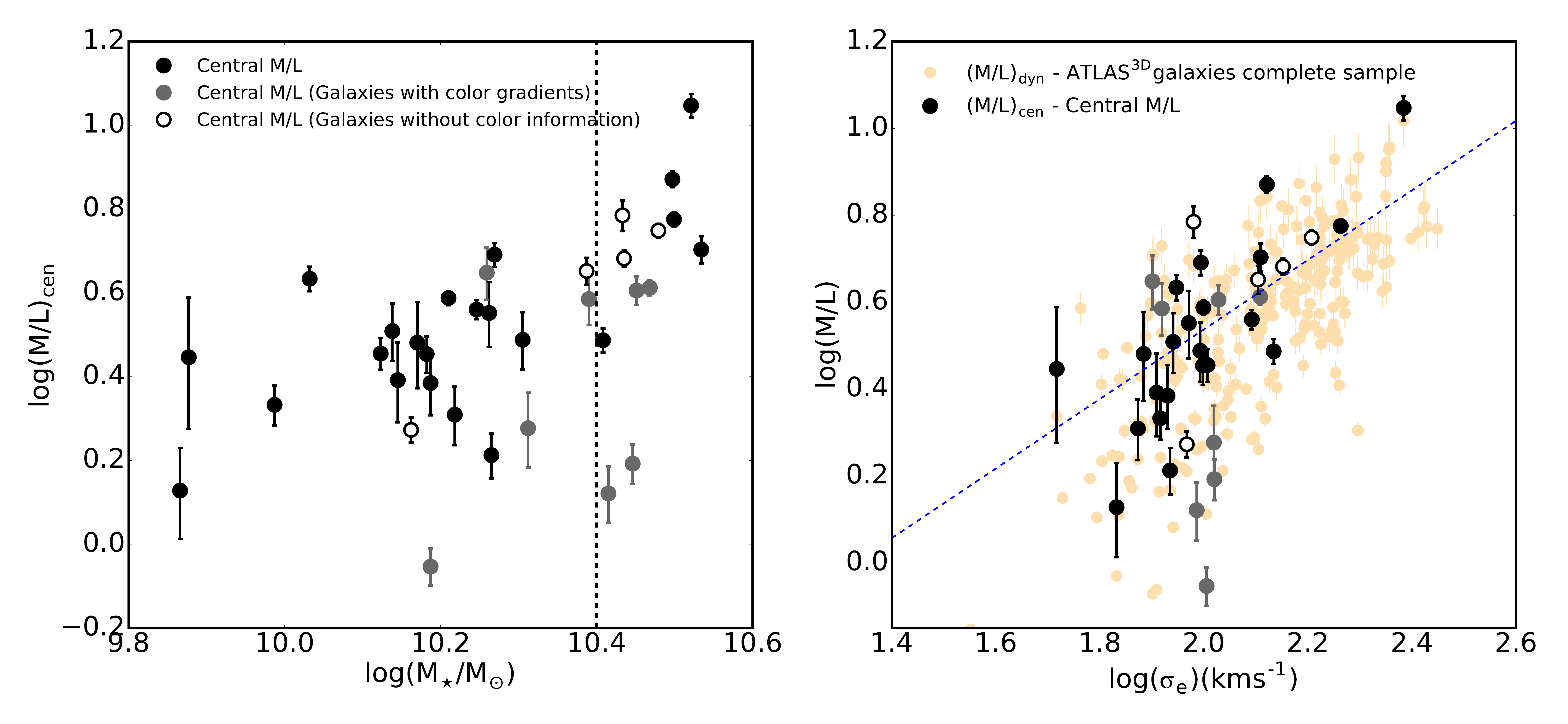}
\caption {\textit{Left}: Best-fit central $(M/L)_{\rm cen}$ in $r$-band derived from JAM modeling of the galaxy sample for a no-BH model, plotted against stellar dynamical mass of the galaxy. The error bars are from the errors on central velocity dispersion translated into $M/L$ errors. Galaxies in gray have color gradients within the central $1''$. Galaxies with open circles do not have color information. \textit{Right}: Best-fit central $(M/L)_{\rm cen}$ in $r$-band (in black, gray and open circles) and $(M/L)_{\rm dyn}$ (in orange) from \citet{atlas3d20} which were derived using a similar method are plotted against effective velocity dispersion. Blue dashed line gives the global relation from Figure 17 of \citet{cappellari16}.}
\label{fig:6}

\end{figure*}
\subsection{Determining central M/Ls with Jeans Anisotropic Models}
\label{sec:jam}
To create a dynamical model for comparing the central velocity dispersion, we use Jeans Anisotropic Models\footnote{\tt Available from http://purl.org/cappellari/software} \citep[python version JAM;][]{cappellari08}. As an input, the models require a mass model parameterized as an MGE. In brief, the JAM code constructs a three-dimensional (3D) density profile by deprojecting the 2D MGE parameters. A gravitational potential is then generated based on the density profile. A BH can be added to this potential, and is represented as a small Gaussian component. Even though a Keplerian potential can be used to describe the BH, it is easier and more efficient to model it as a Gaussian with a width smaller than those used for the galaxy components \citep{emsellem94} (FWHM of BH/FWHM of PSF $\sim$0.015). The Jeans' equations are then solved using the MGE models and the predicted model kinematics are analytically integrated along the line of sight. They are then convolved with the PSF for comparison to the observed galaxy kinematics.

\par To obtain the best-fit central dynamical $M/L$ \{$(M/L)_{\rm cen}$\}, we ran JAM models on the central pixel of the galaxies assuming a no-BH model for our sample of galaxies. We did not include any dark matter component, as the fraction of dark matter at the centers of galaxies is observed to be low (see Section \ref{sec:compareml}). JAM models require a kinematic PSF and a pixel scale as an input. We use a pixel size of $0.94''$ which was the original scale of the SAURON data cube. We derived kinematic PSFs for all the galaxies by comparing the SAURON intensity with that of the model MGE images that were convolved with a circular Gaussian distribution. We then minimized the differences between those images and obtained the PSF width. Based on this method, we found PSFs FWHM ranging from $1''$ to $3''$ with a median of $2''$.

\par To examine the effects of inaccurate PSF determinations, we performed tests by varying the sizes of PSFs from $1''$ to $3''$ FWHM. The change in the derived $(M/L)_{\rm cen}$ due to these PSF variations was found to be around 1\% which is small compared to the errors on the $M/L$ from other sources and in particular small compared to expected BH effect (10-15\%). We also tested whether offsets of the galaxy kinematic data within the central pixel could cause significant errors in our $M/L$ estimates. The median and maximum differences when evaluating our off-center by up to a half pixel were $0.2\%$ and $1.8\%$ respectively, which is again very low compared to our overall errors.

\par Inclination and anisotropies are required as an input to the JAM models. Since, these parameters for a galaxy cannot be found when only a single pixel is modeled, we used the anisotropies\footnote{\tt Available from http://purl.org/atlas3d} and inclinations for the galaxies in our models from Table 1 of \citet{atlas3d15}, which were derived using the full Field of View (FOV) excluding the central $2''$.

\par To compare all the $(M/L)_{\rm cen}$ in a single filter, and for consistency with previous ATLAS$^{3D}$ measurements, we converted all our derived $M/L$ values to SDSS $r$-band values.  This conversion was done in two ways: (1) For 22 galaxies with $HST$ color information, we used the $HST$ colors and magnitudes to estimate the $r$-band luminosity using conversions from Padova SSP models assuming solar metallicity \citep{bressan12}\footnote{\tt Available from http://stev.oapd.inaf.it/cgi-bin/cmd}. (2) For 5 galaxies that did not have color information, 10 Gyr SSP models were used to convert the HST magnitudes to $r$-band luminosity. Note that these converted HST fluxes lie within 1\% of the fluxes from SDSS $r$-band images.
\par A plot of derived $(M/L)_{\rm cen}$ is shown in Figure \ref{fig:6}. The left panel shows $(M/L)_{\rm cen}$ plotted against the total stellar galaxy mass ($M_\star$; defined in Figure~\ref{fig:1}).
The right panel shows the $(M/L)_{\rm cen}$ plotted against the effective stellar velocity dispersion of the galaxy. The $(M/L)_{\rm dyn}$ (in orange; fitted within $\sim$1$R_e$) are taken from \citet{cappellari16}.  The dashed line shows the \citet{cappellari16} global relation between $r$-band $(M/L)_{\rm dyn}$ and the velocity dispersion within one effective radius, $\sigma_e$ of the galaxy: $(M/L)_{\rm dyn} = 6 \times (\sigma_e / 200 {\rm km \ s^{-1}})$.  

The distances and extinction values that are required for the JAM models were obtained from Table 3 of \citet{atlas3d1}. The distances to these galaxies range from 11~Mpc to 41~Mpc with a median of 16.5~Mpc. The extinction values are in $A_b$ mag. We converted all of these values into corresponding values in various filters by using extinction ratios from the Padova model website.

\subsection{Sources of Error on $(M/L)_{\rm cen}$}
\label{sec:errors}
In this section, we look at different sources of error in deriving our $(M/L)_{\rm cen}$ values, including both intrinsic and systematic errors.  Above, we have already considered several sources of error (miscentering of the galaxy, misestimation of the SAURON PSF etc.), and found these to be at the $\sim$1\% level. Here, we consider the two main sources of error in the $(M/L)_{\rm cen}$ estimates of our sample galaxies. 
\begin{enumerate}
\item Uncertainties in the central velocity dispersion of galaxies from SAURON data.
\item Systematic errors that include instrumental effects, PSF effects and errors from JAM modeling.
\end{enumerate}

\par The uncertainties in central velocity dispersion are easy to handle as they can be directly included in modeling to derive the error limits on $(M/L)_{\rm cen}$. These errors, which form the major constituent of the errors in our data, range from 2-9 km/s on the central velocity dispersion of the galaxies, corresponding to fractional dispersion errors from 2\% - 15\%. We calculate the $1\sigma$ error in $M/L$ by adding and subtracting the $1\sigma$ error on central velocity dispersion in our JAM models; this is then translated to an error on the best-fit $M/L$. The errors on our inferred log$(M/L)_{\rm cen}$, from the velocity dispersion errors, range from 0.012 dex - 0.14 dex (3\% - 38\%).

\par
\begin{figure}[ht]

\epsscale{1.25}
\plotone{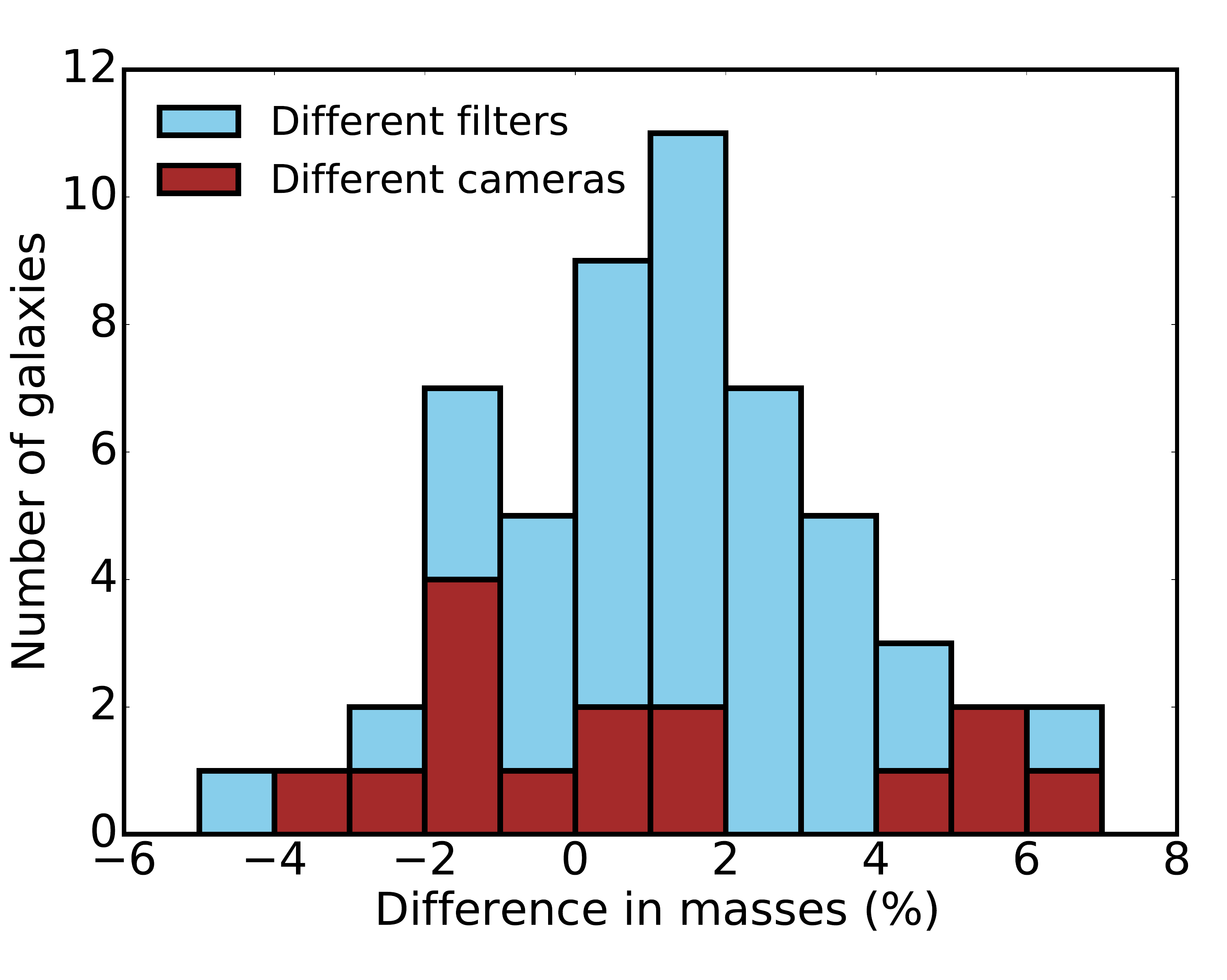}
\caption {A histogram of systematic errors due to uncertainties in the mass models. The masses for each galaxy within the central 1$''$ are calculated using mass models created from two different filters (on the same camera; shown in blue) or two different cameras (shown in red).  The total histogram is best fit by a Gaussian with a standard deviation of 2.53\% and a mean of 1.09\%; we add this error to the errors due to dispersion in calculating our $(M/L)_{\rm cen}$ errors.}
\label{fig:7}
\end{figure}

 For our $(M/L)_{\rm cen}$ measurements, the dominant systematic error comes from errors in our mass models. We derive these errors by determining $(M/L)_{\rm cen}$ for a galaxy (using the method described in Section~\ref{sec:mges} and \ref{sec:dynamicalmodeling}) in two different $HST$ filters observed with the same camera and then calculate the difference in mass derived in the central arcsec from the two filters. We also compare independent mass models derived from data in two different cameras, but with similar filters. For an object with a uniform color (i.e.~after excluding the objects with dust lanes or significant color gradients), the mass models of a galaxy derived from two different filters should be identical. Those in two different cameras with the same filter, mass models should be identical regardless of the sample. However, misestimates in the PSF, residuals in the modeling, and random sampling errors of the surface brightness profile can lead to differences in these mass models. We can characterize the size of the systematic errors due to these by comparing the two filter and two camera models. The differences in these models is shown in Figure~\ref{fig:7}. We represent this model as a Gaussian error with standard deviation of 2.5\% to match the width of this distribution and consider it as the systematic error throughout the paper.

To summarize, our major sources of errors on our $(M/L)_{\rm cen}$ estimates are:
\begin{itemize}
\item Uncertainties from central velocity dispersion: 3-38\%
\item Systematic mass-modeling errors: 2.5\%
\end{itemize}

\begin{figure*}
\includegraphics[width=\linewidth]{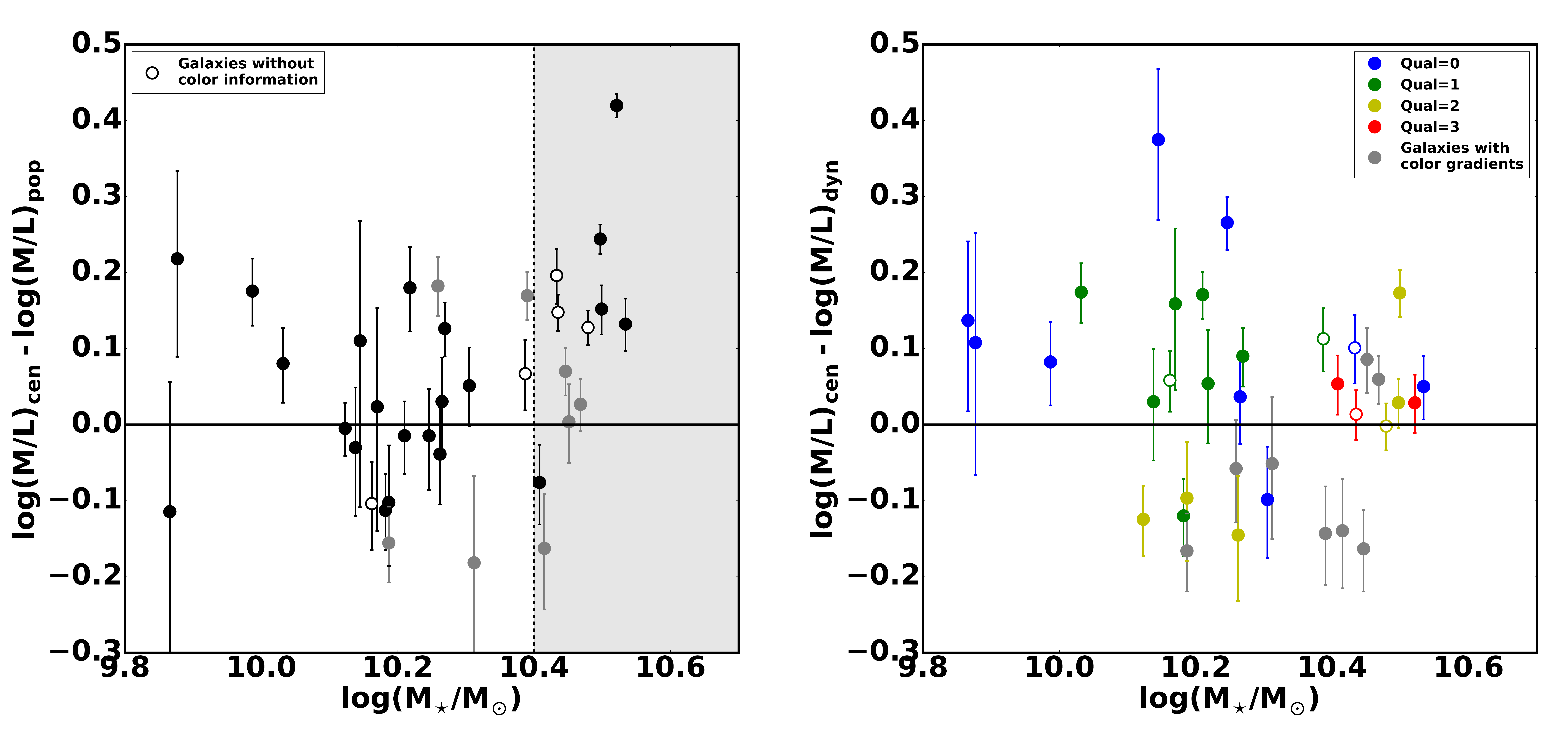}
\caption{The central $M/L$s of low-mass ETGs are typically higher than expected.  \textit{Left Panel:} The ratio of central $(M/L)_{\rm cen}$ and $(M/L)_{\rm pop}$  in $r$-band (derived from stellar population models assuming a Kroupa IMF) plotted against the stellar mass of the galaxy. Galaxies above log$M_\star$ of 10.4 (gray region) appear to have systematically higher $(M/L)_{\rm cen}$ relative to $(M/L)_{\rm pop}$, perhaps due to global IMF variations. \textit{Right Panel:} The ratio of central $(M/L)_{\rm cen}$ and $(M/L)_{\rm dyn}$  in $r$-band (derived from dynamical models fit outside the nucleus and within $\sim$1$R_e$) are plotted against the mass of the galaxy. The galaxies are colored according to the quality of the fit of models for deriving $(M/L)_{\rm dyn}$ \citep{atlas3d15}. Qual=0 indicates a low-quality JAM fit that is due to barred galaxies, dust or uncertain deprojection due to a low inclination of the galaxy. Qual=3 indicates a good JAM fit to the data. The galaxies that had a color gradient in central $0.94''$ are indicated in gray as their $(M/L)_{\rm cen}$ values are likely less reliable. The error bars include both errors from velocity dispersion measurements and mass modeling errors.}
\label{fig:8}
\end{figure*}

\begin{figure*}[ht]
\epsscale{1}
\includegraphics[width=\linewidth]{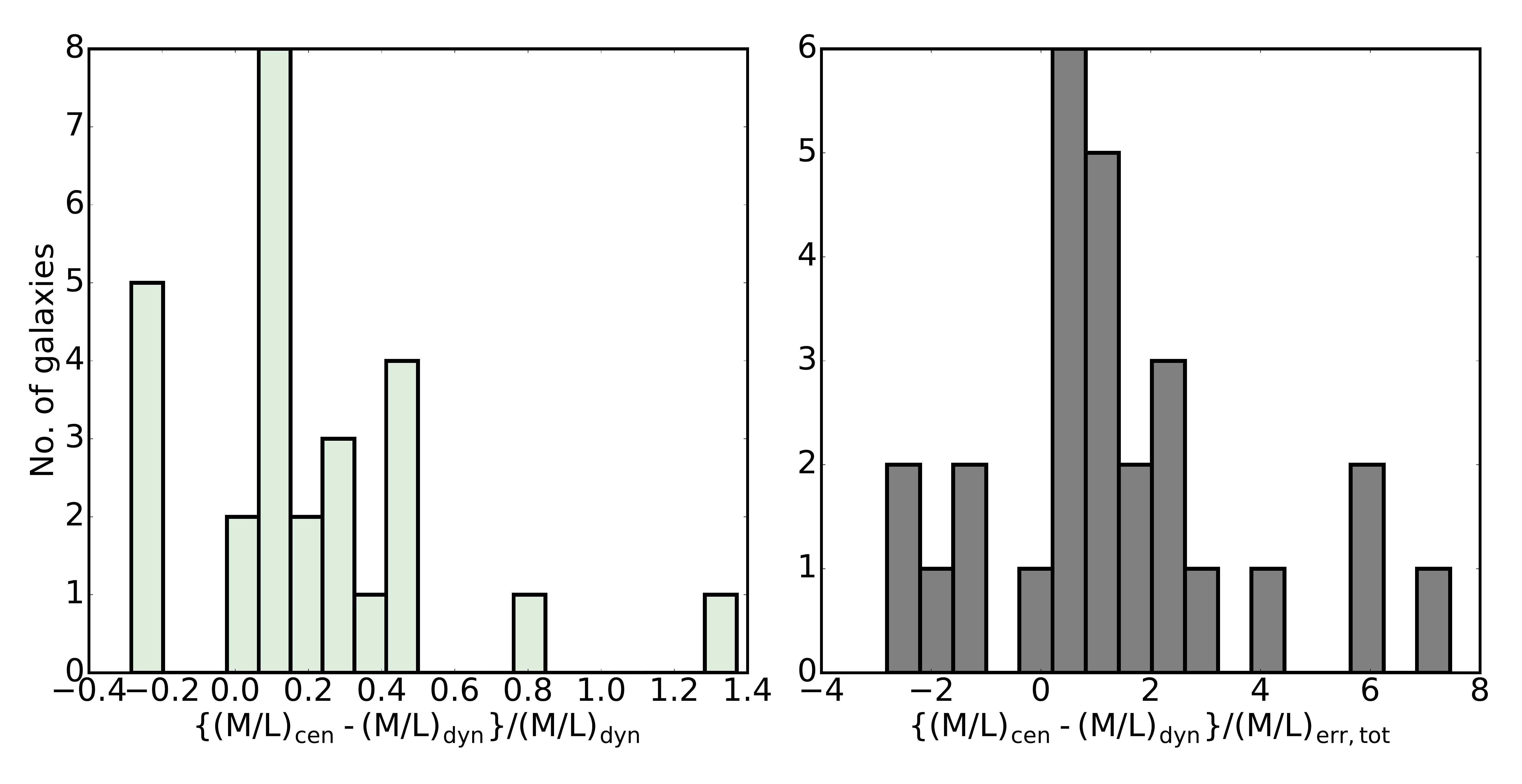}
\caption {\textit{Left Panel:} Distribution of enhancement of central $(M/L)_{\rm cen}$ with respect to $(M/L)_{\rm dyn}$. This indicates the percentage by which $(M/L)_{\rm cen}$ is higher than $(M/L)_{\rm dyn}$; the median enhancement is 14\%.  \textit{Right Panel:} Distribution of the significance of central $(M/L)_{\rm cen}$ of our sample of galaxies. The plot indicates the significance in the enhancement of central $M/L$. Four high significance detections are at the right end. }
\label{fig:9}

\end{figure*}

\section{Results: Enhancement in the central dynamical $M/L$}
\label{sec:compareml}
In the previous sections, we described the derivation of our mass models and corresponding determination of central, dynamical $(M/L)_{\rm cen}$ for our 27 low-mass ETGs. In the following, we compare the derived $(M/L)_{\rm cen}$ for the HST images with the mass-to-light ratios derived from stellar populations estimates of the nuclear spectra \{$(M/L)_{\rm pop}$\} and dynamical mass-to-light ratio derived outside the nucleus \{$(M/L)_{\rm dyn}$\}.  

\par As discussed in Section~\ref{sec:inner_outer} the $(M/L)_{\rm pop}$ was derived by fitting the central ATLAS$^{3D}$ spectrum to SSP models \citep{atlas3d30}. For the comparison to $(M/L)_{\rm cen}$ we use fits done with a Kroupa IMF, as these lower mass ETGs are expected to have an IMF similar to Kroupa \citep{cappellari12}. The $(M/L)_{\rm dyn}$ was derived by dynamically modeling the 2-D ATLAS$^{3D}$ kinematics at radii beyond $2''$ using JAM models \citep{atlas3d15,atlas3d20}. These dynamical models exclude the central regions ($< 2''$) of the galaxy due to a possible bias from super massive black holes being at their centers. The stellar and dark matter components were fit separately in the models \citep[Model B of ][]{atlas3d15}. For the $(M/L)_{\rm dyn}$, the best-fit models were used to predict the total dynamical mass at the center, including adding in the best-fit dark matter profile. This dark matter component makes up only 2-3\% of the central mass in our sample galaxies.  The $(M/L)_{\rm dyn}$ is therefore the prediction for what we should observe for $(M/L)_{\rm cen}$ based on the dynamical model fits to the outer parts of each galaxy.

\par A comparison of $(M/L)_{\rm pop}$ and $(M/L)_{\rm dyn}$ with $(M/L)_{\rm cen}$ is shown in Figure~\ref{fig:8} plotted against the galaxy stellar mass. We find that $(M/L)_{\rm cen}$ is larger than $(M/L)_{\rm dyn}$ in 21 of 27 galaxies, and larger than $(M/L)_{\rm pop}$ in 17 of 27 galaxies. A trend is also seen in the left panel for galaxies that are massive (log($M_{\star}/M_{\odot})>10.4$). Their high mass-to-light ratios as compared to $(M/L)_{\rm pop}$ (but not $(M/L)_{\rm dyn}$) may be due to systematic IMF variations observed using similar dynamical approach in ETGs \citep{cappellari12,atlas3d20,posacki15} and in the general galaxy population \citep{Li17}. The galaxies with significant central color gradients (which can result in biased values of $(M/L)_{\rm cen}$) are shown in gray in these plots, but excluded from our analysis hereafter.   

\par Due to the offset from stellar populations values at the higher mass end, we focus on the comparison between $(M/L)_{\rm cen}$ and $(M/L)_{\rm dyn}$ as shown in the right panel of Figure~\ref{fig:8}. We highlight the significance of the differences between $(M/L)_{\rm cen}$ and $(M/L)_{\rm dyn}$ in Figure~\ref{fig:9}; both panels show the clear offset towards higher values for $(M/L)_{\rm cen}$ relative to $(M/L)_{\rm dyn}$. The left panel shows a median offset of 14\%. Assuming a Gaussian distribution, we calculate that the probability of randomly having 21 of the 27 $(M/L)_{\rm cen}$ values higher than the $(M/L)_{\rm dyn}$ values is just 0.002, therefore the enhancement is statistically significant at the $>$3$\sigma$ level based simply on the asymmetry of the distribution.

\par The right panel of Figure~\ref{fig:9} shows the differences relative to the expected errors, showing that some individual objects have highly significant enhancements in $(M/L)_{\rm cen}$. $(M/L)_{\rm err,tot}$ is the total $1\sigma$ error that has several sources of error added in quadrature:
\begin{equation}(M/L)_{\rm err,tot}^2 = (M/L)_{\rm err,rms}^2 + (M/L)_{\rm err,sys}^2 + (M/L)_{\rm err,dyn}^2\end{equation} 
where $(M/L)_{\rm err,rms}$ (2-15\%) is error from the measurements of velocity dispersion and $ML_{\rm err,sys}$ (2.5\%) is the systematic error (both described in Section~\ref{sec:errors}). The ($M/L)_{\rm err,dyn}$ is the error on $(M/L)_{\rm dyn}$; for this error we take the quoted 6\% uncertainty given in \citet{atlas3d15}.  We note that same uncertainty is assumed for both $(M/L)_{\rm err,sys}$ and $(M/L)_{\rm err,dyn}$ in all galaxies.

\section{Causes of Inflated $M/L$}
An enhancement in the central $M/L$ is clearly seen in our sample of galaxies. We have already excluded several systematic effects that could cause this positive bias in the previous sections. Here we discuss additional possibilities.

\par We start by considering whether the enhanced $M/L$ could be due to gradients in stellar ages and metallicities. Because $(M/L)_{\rm dyn}$ is derived from kinematics at larger radii, any gradient in stellar populations between those radii and the center could lead to differences in the stellar $M/L$. But based on the stellar population fits to the ATLAS$^{3D}$ spectra discussed in Section~\ref{sec:inner_outer} and shown in Figure~\ref{fig:5}, the central $(M/L)_{\rm pop}$ are on average lower than the outer $(M/L)_{\rm pop}^{\rm out}$. The maximum enhancement in the $(M/L)_{\rm cen}$ in Figure~\ref{fig:5} is $\sim$$7\%$, which is much lower than the values in Figure~\ref{fig:9}, which have a median enhancement of $\sim$14\%. Thus, based on stellar populations variations alone, we'd expect a {\em decline} in the central $M/L$ for most galaxies, not the increase that is observed.  Including these variations in stellar population $M/L$ as a correction to the $(M/L)_{\rm cen}$ comparison to $(M/L)_{\rm dyn}$, we find enhanced $(M/L)_{\rm cen}$ in 21 of 27 galaxies, with a median enhancement of 18\%, very similar to the values found without correction. This strongly suggests a metallicity or stellar age gradient cannot be the reason for an enhancement in the $M/L$.   
\par These stellar population gradients may provide an  explanation for the negative tail of the distribution extending to higher significance than expected based on our error estimates.   This could happen if metallicity or stellar age gradients lead to an underestimation of the central $(M/L)_{\rm cen}$. We find that five out of six galaxies with decreased central $M/L$s have a younger population at their centers.

\par Overall, this leaves us with three possible explanations for the enhanced $(M/L)_{\rm cen}$ values:
\begin{enumerate}
\item The presence of massive BHs in these galaxies. 
\item Radial IMF variations in the galaxy.
\item Variations in Anisotropy
\end{enumerate}
We focus on these three explanations in the next sections. If we assume the $(M/L)_{\rm cen}$ enhancements are due to the presence of central BHs, we can derive BH masses for these galaxies, and compare these to previous dynamical estimates.  We also constrain the scatter in the $M-\sigma$ relationship using the enhancement in the central $M/L$. Then in the following subsections, we discuss the possibility of radial IMF variations and the effects of anisotropy in determination of $M/L$. 

\subsection{Massive Black Holes}

\begin{figure*}[ht]
\epsscale{1}
\label{fig:10}
\includegraphics[width=\linewidth]{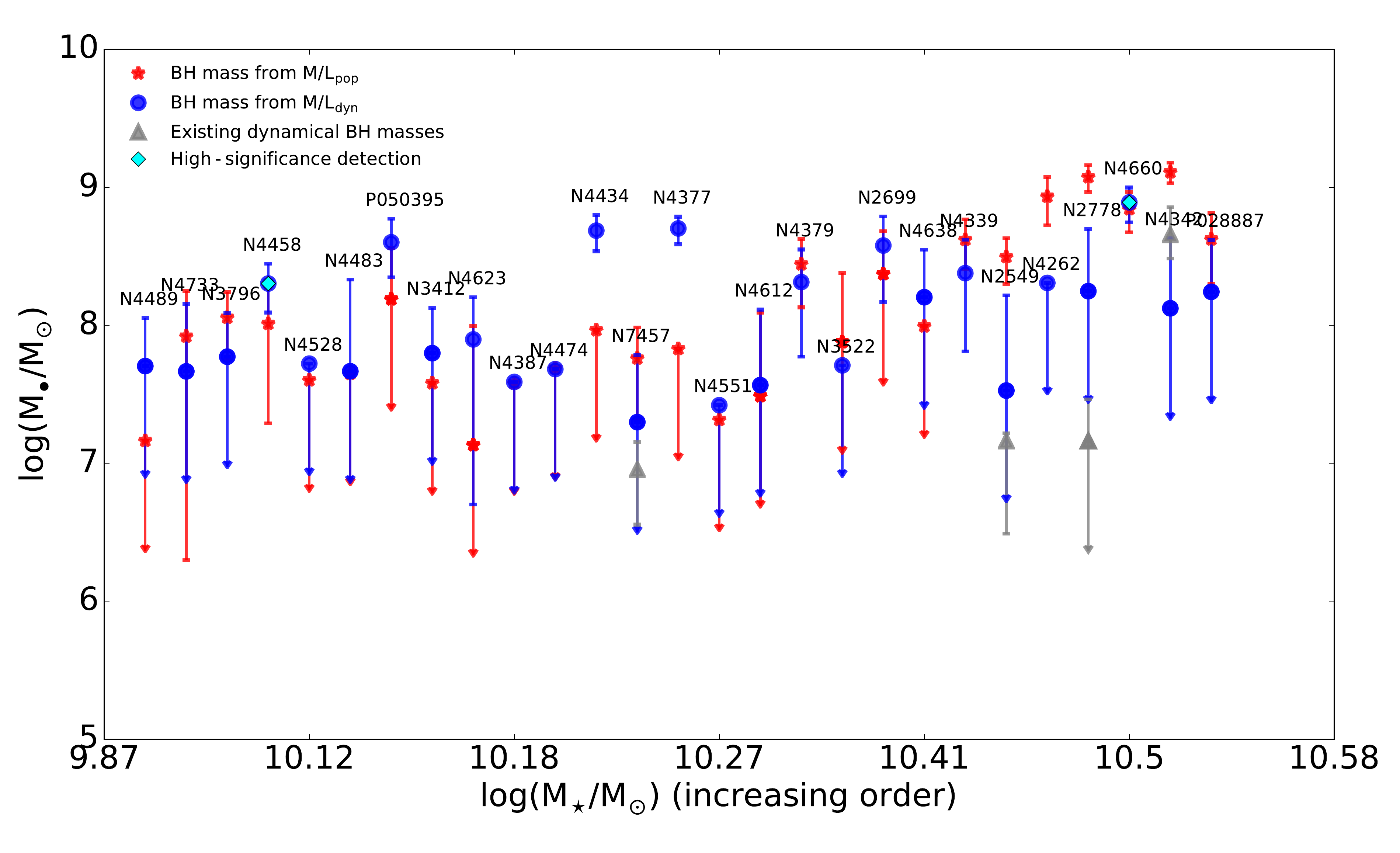}
\caption {Estimated BH masses and their upper limits for galaxies calculated using JAM models based on the observed enhancement in the $(M/L)_{\rm cen}$. These are plotted from lowest to highest stellar masses. The BH masses calculated using $(M/L)_{\rm dyn}$ are plotted in blue whereas BH masses calculated using $(M/L)_{\rm pop}$ are plotted in red. Previously known BH masses are plotted in gray.  We highlight the two highest significance detections as cyan diamonds.}
\includegraphics[width=\linewidth]{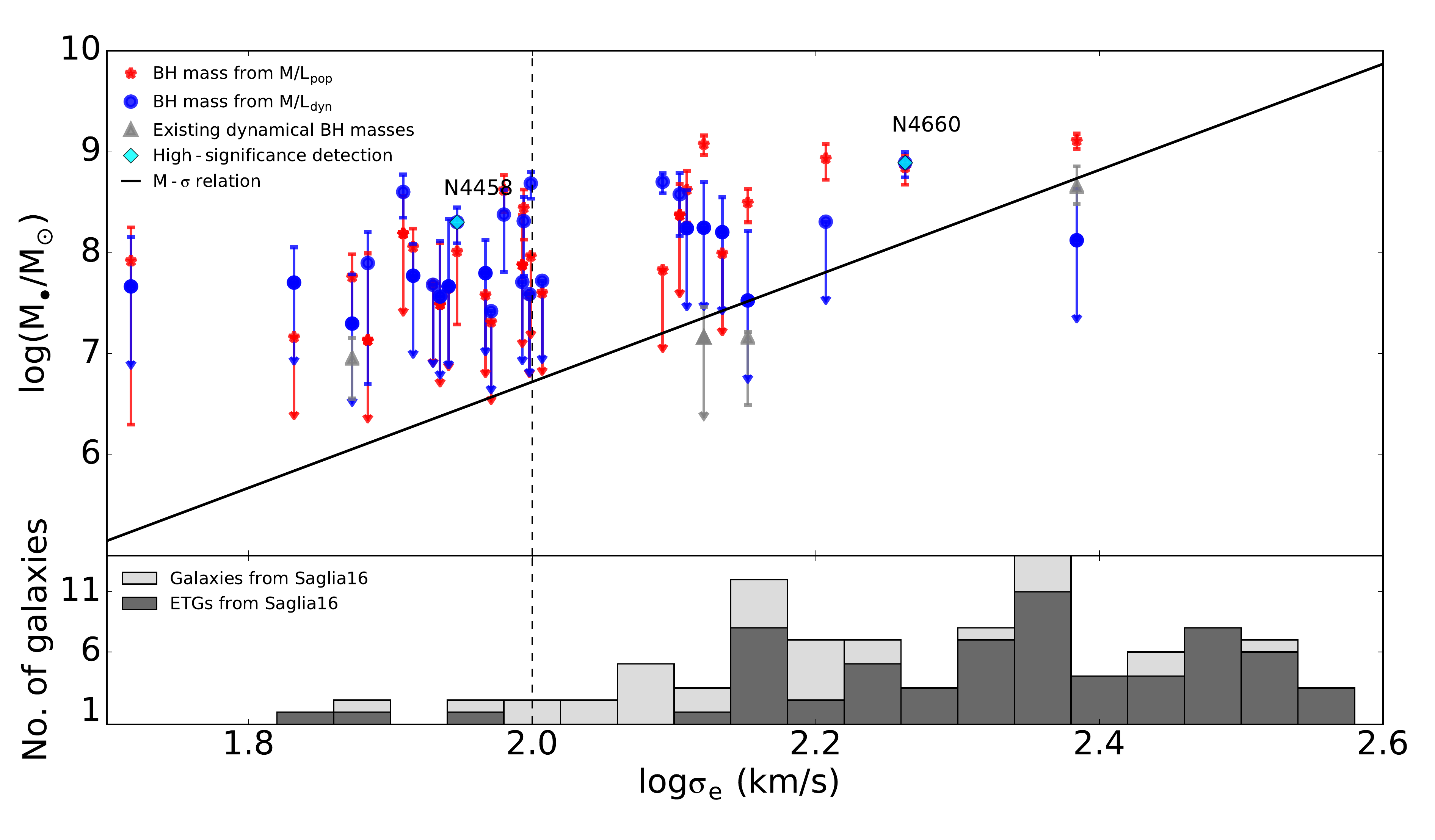}
\caption {\textit{Top Panel}:Estimated BH masses plotted in conjunction with the M$_\bullet$$-\sigma$ relationship (solid line).  Our method is sensitive only to BHs with masses higher than expected from the M$_\bullet$-$\sigma$ relationship. Symbols as in Figure~\ref{fig:10}. \textit{Bottom Panel}: The distribution of velocity dispersion of galaxies from \citet{saglia16}. Note that there are very few ETGs having $\sigma<100$ km/s and we are probing this low velocity dispersion region. }
\label{fig:11}
\end{figure*}

\par The nuclei's elevated $M/L$ may be due to the presence of a BH; if this were the case, we could use these to estimate the BH mass required to account for the apparent change in $M/L$. This approach is essentially identical to that of \citet{mieske13} who suggested BHs as the explanation for enhanced $M/L$s in UCDs. Estimates of BH masses for each galaxy were derived using JAM models based on the central velocity dispersion and taking $(M/L)_{\rm dyn}$ (calculated outside the nucleus) as the true stellar $M/L$. The amount by which $(M/L)_{\rm cen}$ is higher than $(M/L)_{\rm dyn}$, therefore directly translates to a best-fit BH mass. We also derived BH masses considering the $(M/L)_{\rm pop}$ values as the true stellar $M/L$. The errors on our BH mass estimates are directly translated from the errors as described in Section~\ref{sec:errors}. We also get the upper limits for the galaxies' BH masses, where $(M/L)_{\rm cen} < (M/L)_{\rm dyn}$, by considering the minimum limit to be $(M/L)_{\rm dyn}$ and adding the $1\sigma$ error to it. Figure~\ref{fig:10} shows our calculated BH masses based on $(M/L)_{\rm dyn}$ and $(M/L)_{\rm pop}$, ranked in increasing order of their galaxy mass.  Using $(M/L)_{\rm dyn}$, we find 9 galaxies $>$$1\sigma$ enhancements; these galaxies have BH mass estimates ranging from $\sim$8$\times$$10^7$~M$_\odot$ to $\sim$8$\times$$10^8$~M$_\odot$. Our upper limits are all above $\sim$10$^7$~M$_\odot$, suggesting we don't have a sensitivity to BHs below this mass limit using this technique.  The estimated black hole masses, $M-\sigma$ relation predictions and the sizes of the central regions for each galaxy are shown in Table~\ref{table:bhmass}.

In Figure~\ref{fig:11}, we show the same BH mass estiamtes, but plotted against galaxy velocity dispersion $\sigma_e$.  The bottom histogram shows that our sample falls below the dispersion of most previous dynamical BH mass estimates.  All of our estimated BH masses lie clearly above the $M-\sigma$ relationship.


We stress that the BH masses inferred here are crude estimates and our upper limits are very dependent on the determination of our errors. Nonetheless, the UCD case shows that the $\sim$10\% mass fraction BHs that were suggested by the similar \citet{mieske13} analysis have in fact been found using high-resolution spectroscopy \citep{seth14,ahn17}, and the size of the nuclear $M/L$ enhancements we see in the lower mass ATLAS$^{3D}$ ETGs are comparable to the level of enhancements we expect if similar black holes exist in these galaxies (see Section~\ref{sec:ucds}).


Using the estimated BH masses, we calculate the sphere of influence (SOI) radius, $r_{SOI}~=~G.M_{\bullet}$/$\sigma{_e^2}$, listed in Table~\ref{table:bhmass}. The median $r_{SOI}$ of 38~pc is comparable in size to the median pixel size of our sample galaxies of 75~pc (also listed in Table~\ref{table:bhmass}).  However, the $r_{SOI}$ is also significantly smaller than the central pixel size in some cases, as expected when the stellar component dominates the kinematics of the central pixel; this does not in principle prevent us from correctly estimating the BH mass.  We note that the mass models of our galaxies have a median pixel size (from HST imaging) of 4~pc; variations in the mass model on scales smaller than this could lead to systematic errors in our BH mass estimates.

We highlight particularly interesting objects here.

Two galaxies NGC~4458 and NGC~4660 show $\gtrsim$3$\sigma$ enhancements in $(M/L)_{\rm cen}$ relative to both the $(M/L)_{\rm dyn}$ and $(M/L)_{\rm pop}$ values.  Both have robust mass models, are not affected by significant central color gradients, and are therefore our strongest candidates for BH detections.  In both cases, the BHs inferred are much more massive than the prediction from the $M-\sigma$ relation (70 times and 6 times respectively).  NGC~4458 is a relatively low-mass ETG ($M_\star = 1.07\times10^{10}M_{\odot}$), while NGC~4660 is about three times more massive.  

Two other objects, NGC~4377 and NGC~4434 also have a $>$3$\sigma$ enhancement of $(M/L)_{\rm cen}$ relative to $(M/L)_{\rm dyn}$, but are not at all enhanced relative to $(M/L)_{\rm pop}$.  This could be due to errors in the low-quality \citet{atlas3d20} dynamical models (quality values of 0 and 1 respectively), and we do not consider either of these as robust BH candidates. An additional nine objects, mostly at the high mass end, have $>$3$\sigma$ enhancements in $(M/L)_{\rm pop}$, but $<$3$\sigma$ enhancements in $(M/L)_{\rm dyn}$; of the NGC~4339 and NGC~4379 have the most significant $(M/L)_{\rm dyn}$ enhancements and are most likely to be BH detections. However, for the bulk of these objects, we suspect that they have an IMF heavier than the Kroupa IMF as suggested by their high $(M/L)_{\rm dyn}$ values at larger radii \citep{cappellari12,atlas3d20}. To recap, we have enhancement in central $M/L$, $>$ 1$\sigma$ for 14 galaxies,$>$~2$\sigma$ for 8 galaxies and $>$~3$\sigma$ for 4 galaxies with respect to $(M/L)_{\rm dyn}$.

\subsubsection{Comparing BH masses with existing BH masses}
\par Five galaxies in our sample have existing dynamical measurements of BH masses or upper limits \citep[as compiled][]{kormendyandho13}. These are also plotted in Figure~\ref{fig:10} (in gray); NGC~2549, NGC~2778, NGC~3377, NGC~4342 and NGC~7457. Note that the existing measurement in NGC~2549 is a weak detection \citep{krajnovic09} and NGC~2778 only has an BH mass upper limit \citep{schulze11}.

For NGC~3377 (not shown due to its color gradient), we get a $(M/L)_{\rm cen}$ of 4.09, which compared to the $(M/L)_{\rm dyn}$ of 3.57 suggests an $M_{BH} =$9$\times$10$^7$~M$_\odot$ which is consistent with its existing BH mass. In the other four galaxies, all the BH mass estimates are consistent with our BH upper limits determined from $(M/L)_{\rm dyn}$. However, many of them have much higher inferred BH masses from $(M/L)_{\rm pop}$, we take this as further evidence that our enhancements in $(M/L)_{\rm cen}$ relative to $(M/L)_{\rm pop}$ values are not due to BHs, at least for galaxies above $\sim$10$^{10.4}$~M$_\odot$. Instead, these enhancements in dynamical vs.~population $M/L$s both in the center and outer regions suggest global IMF variations. As noted in the introduction, the sample of existing BH masses in this mass range of ETGs is minimal, and none of our strongest BH candidates have previously published dynamical BH mass estimates. However, there are existing central X-ray measurements for NGC~4660 (4.46$\times$$10^{38}$ ergs s$^{-1}$), NGC~4379 (4.16$\times$$10^{38}$ ergs s$^{-1}$) and NGC~4458 ($<$ 2.13$\times$$10^{38}$ ergs s$^{-1}$)\citep{gallo10}.
 \par To further check the accuracy of our measured BH masses, we considered a sample of 10 high-mass ETGs ( log(M$_\star$/M$_{\odot}) > 10.5$ ) from the ATLAS$^{3D}$ sample that had existing dynamical BH mass estimates from \citep{saglia16}. We modeled those galaxies in a similar way using JAM models as we did earlier, and estimated the BH masses. Eight out ten of these galaxies had BH masses within the $1\sigma$ range of their existing BH measurements.

\subsubsection{Constraining the scatter in $M_\bullet-\sigma$ relationship}
\par We now describe how we can use our existing measurements of the enhancement of $(M/L)_{\rm cen}$ to $(M/L)_{\rm dyn}$ to constrain the scatter in the $M_\bullet-\sigma$ relationship. A large scatter in this mass range of galaxies would result in some of our galaxies having unusually massive BHs that would be detectable as enhancements in $(M/L)_{\rm cen}$.

We first start by predicting the BH masses in each galaxy assuming the $M_\bullet-\sigma$ relationship from \citet{saglia16}:
\begin{equation}\label{msigma} \log M_{\bullet}=5.246\times \log\sigma_e-3.77\end{equation}
The $\sigma_e$ used here is the effective stellar velocity dispersion adopted from Table 1 \citep{atlas3d15}. They were calculated by co-adding all SAURON spectra within an effective ellipse of area $\pi R_e^2$, where $R_e$ is the effective radius. The expected $M_\bullet$ over the range of $\sigma_e$ values of our galaxies are shown as a line in Figure~\ref{fig:11}, and are given for each galaxy in Table~\ref{table:bhmass}. Our BH mass estimates are not sensitive to any BHs that fall on the $M_\bullet-\sigma$ relationship, thus we will only detect BHs when they are significantly more massive than expected. We can therefore constrain the scatter in the $M_\bullet-\sigma$ relationship through our most significant detections. 
\begin{figure}
\centering
\begin{minipage}[b]{.25\linewidth}
  \includegraphics[width=75mm]{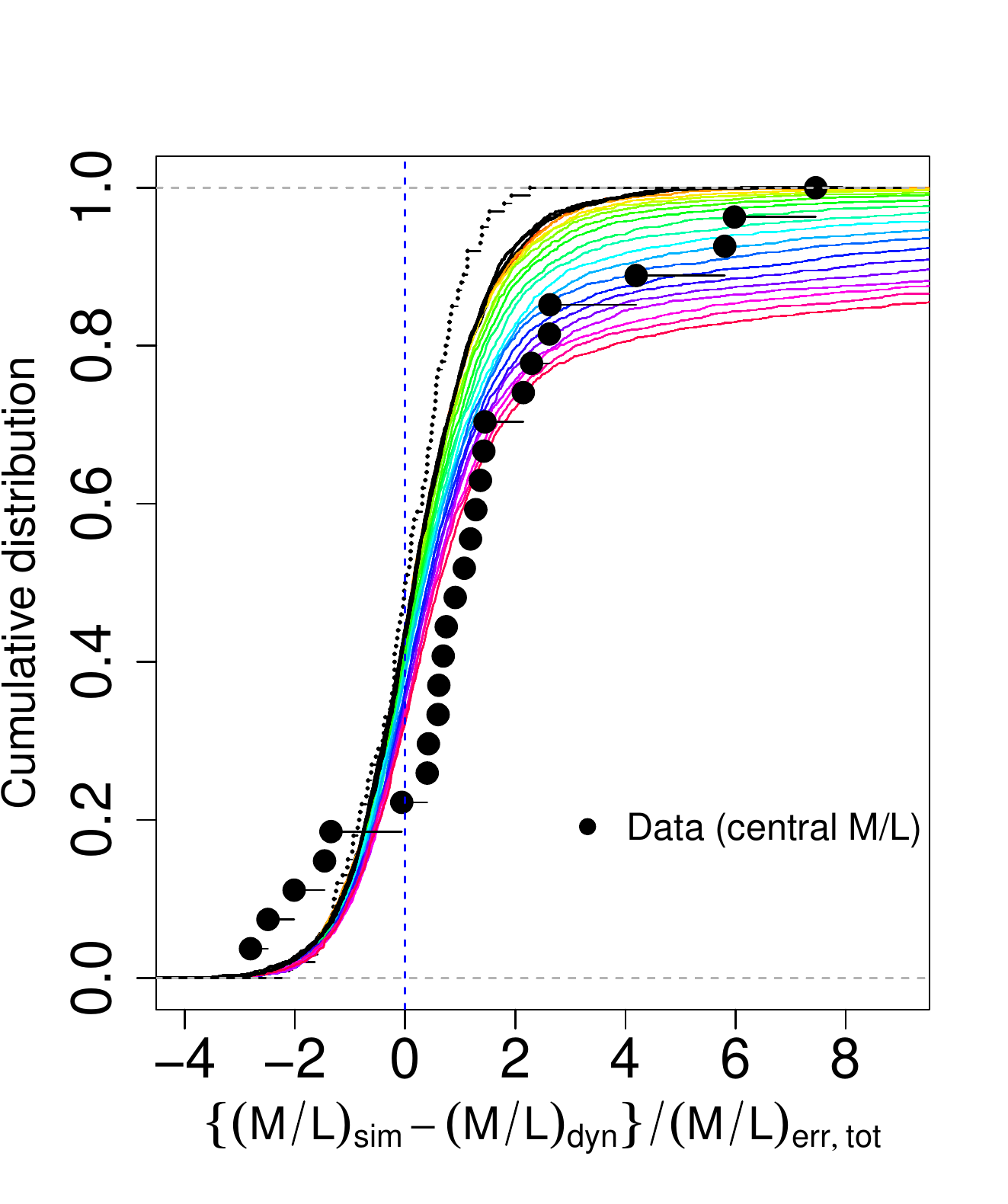}
  
  \label{fig:12}
\end{minipage}
\hspace{.265\linewidth}
\begin{minipage}[b]{.45\linewidth}
  \includegraphics[width=65.3mm]{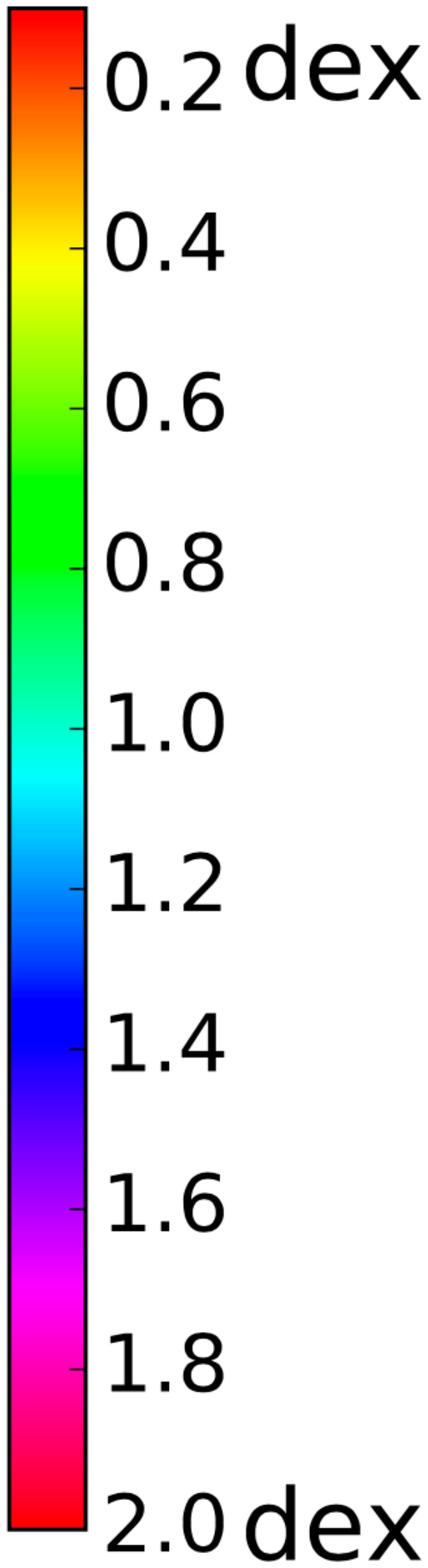}
  \par\vspace{10pt}
\end{minipage}
  \caption{Results from our simulations to constrain the scatter in the $M_\bullet-\sigma$ relationship.  Comparison of the empirical cumulative distribution functions (ECDFs) of the enhancement in central $M/L$ between simulations and data. The solid points are the ECDF of the original data (right panel of Figure~\ref{fig:9}). The dashed black line shows a simulation with no BH including only the errors and is therefore a Gaussian distribution; the solid black line shows the effect of adding $M_\bullet-\sigma$ relationship BHs to each galaxy in our sample. Colored lines show simulations with a range of scatter in the $M_\bullet-\sigma$ relationship. The scatter increases from top to bottom ranging from 0.1 - 2.0 dex.}
\end{figure}

\par Simulations are performed to predict the central $M/L$ of the galaxies from the $M_\bullet-\sigma$ relationship with scatter. Specifically, we use {}the $M_\bullet-\sigma$ relationship from Eq.~\ref{msigma} and add random scatter to it ranging from 0.0 dex to 2.0 dex. A BH mass is computed based on the $\sigma_e$ of the galaxy with the scatter added in as a Gaussian random variable. Using our luminosity models for each galaxy, we predict the central velocity dispersion by assuming the central $M/L$ of the galaxies is equal to $(M/L)_{\rm dyn}$ and adding the effect of the BH. Then, we simulate our observations by adding a 1$\sigma$ Gaussian random error based on the observed dispersion error from ATLAS$^{3D}$. We then derive an $M/L$ from this dispersion using a JAM model, just as we have for our observed data; we add a 2.5\% Gaussian error to account for the systematic effects of the luminosity models, which yields our final $(M/L)_{\rm sim}$ value. The $(M/L)_{\rm sim}$ value is then compared to the $(M/L)_{\rm dyn}$ to determine the enhancement relative to the error. For each level of scatter in the $M-\sigma$ relationship, we run 1000 simulations per galaxy, and calculate the \{$(M/L)_{\rm sim}-(M/L)_{\rm dyn}$\}$/(M/L)_{\rm err,tot}$ for the full set of galaxies. Empirical cumulative distribution functions (ECDFs) of these simulations are shown as the colored lines in Figure~\ref{fig:12}, and can be compared to the ECDF of the observed data (in solid points), which is the cumulative version of the right panel of Figure~\ref{fig:9}.

\par The difference between the dashed black line (the no BH model, including only measurement errors) and the solid black line (a model with $M_\bullet-\sigma$ relation black holes with no scatter) shows that despite being far below our mass sensitivity for individual detections, the presence of $M_\bullet-\sigma$ black holes does have a clear effect on the expected ECDFs, increasing the number of enhanced $M/L$ objects we'd expect.

\par Two trends are apparent comparing the observed and model ECDFs. First, the central $M/L$ enhancement of 0 to 2$\sigma$ do not seem to be consistent with being caused by $M_\bullet-\sigma$ relation black holes regardless of the scatter. We think the most likely cause of these enhanced central $M/L$s are radial IMF variations, which we discuss further in the next subsection. Second, the models with $M_\bullet-\sigma$ scatter above $\sim$1 dex would clearly result in detections of more enhanced $(M/L)_{\rm cen}$ values than we observe. Our highest outlier $(M/L)_{\rm cen}$ is enhanced by 7.5$\sigma$ relative to $(M/L)_{\rm dyn}$ and we can use this maximum value to put a limit on the scatter in the $M_\bullet-\sigma$ relation for $\sim$10$^{10}$~M$_\odot$ ETGs.

\par To quantify these constraints, we examine our 1000 simulations and see how often a single simulation would give an $(M/L)_{\rm cen}$ enhancement larger than the highest outlier in our data. We find that when the scatter is sufficiently large ($>$1.2 dex), 95\% of our simulations predict a central $M/L$ greater than 7.5$\sigma$.  On the other hand, when it is below 0.4 dex, only 5\% of the simulations predict it. Thus with 95\% confidence, we can constrain that the scatter in the $M_\bullet-\sigma$ for our low mass ETGs is between 0.4 dex - 1.2 dex.  Recent measurements suggest a scatter in the $M_\bullet-\sigma$ relation of $\sim$0.4 dex for the mostly higher dispersion galaxies with existing dynamical BH mass measurements \citep[e.g.][]{saglia16,greene16}, suggesting the scatter either stays constant or grows in these lower mass ETGs. 

\subsection{Radial IMF variations in the galaxies} 
\par Apart from black holes, another plausible cause for the increase in $(M/L)_{\rm cen}$ relative to $(M/L)_{\rm dyn}$ are radial gradients in the IMF. Evidence for enhancements in the number of low-mass stars have been suggested from both spectroscopic studies of dwarf sensitive lines \citep[e.g.][]{vandokkum10nat,vandokkum11,conroy12,spiniello12,lb17} and dynamically due to higher measured $M/L$s \citep[e.g.][]{cappellari12,tortora13,lyubenova16}. Recent work has explored whether these IMF variations are radially varying within the galaxies, with a mix of results including some evidence for increasingly strong dwarf sensitive line indices in the central regions of galaxies ($\sim$300~pc) \citep{spiniello14,navarro15,mcconnell16}. For example, \citet{navarro15} find that the IMF slope varies from $\alpha=3.0$ at the center to $\sim$1.8 at an effective radius, with most of the change being within 0.2$R_e$.  However, the recent work of \citet{davis17} shows a wide range of behavior in the apparent radial IMF gradients.

The largest variations in IMF appear to occur in the most massive, highest dispersion galaxies, and therefore, most studies on global and radial IMF variations have focused on galaxies with $\sigma > 200$~km/s.  However, our sample is quite different from most of the galaxies in these previous studies of the IMF, with all but one of our galaxies having $\sigma < 200$~km/s. 

We have shown above that the small enhancements in $(M/L)_{\rm cen}$ relative to the $(M/L)_{\rm dyn}$ derived at larger radii are seen in a majority of our galaxies. These enhancements are not easily interpreted as being due to BHs drawn from the $M_\bullet-\sigma$ relation; our measured ECDF in Figure~\ref{fig:12} has many more galaxies at $+$0-2$\sigma$ enhancements than any of the BH models. The typical $(M/L)_{\rm cen}$ enhancements in our galaxies are $\sim$14\% (Figure~\ref{fig:9}), and the $M/L$ variation expected from the inferred IMF variations in massive galaxies can be significantly larger than this \citep{davis17}. Thus, it is plausible that small changes in the IMF slope could be creating the enhanced $M/L$ we observe. 

We tested the amount of change in the IMF slope that would be required to generate an enhancement of 14\% in the inferred $M/L$. We find that a small change in the IMF slope from $\alpha=2.3$ to $\sim$2.4 is enough to create this difference. We also derived the inferred $M/L$ for the galaxies from the BH masses calculated from the $M-\sigma$ relationship. The inclusion of these BHs only slightly lowers the median $M/L$ enhancement to  12\%.  We also note that the stellar population variations shown in Figure~\ref{fig:5} would make this $M/L$ enhancement larger.  In any of these cases, a small change in the IMF slope could be responsible for the $M/L$ enhancements with the IMF being steeper in the center compared to their outer regions.

\subsection{Impact of anisotropy}

\par Differences from the anisotropies assumed in our models may also lead to variations in central $M/L$. Since we model only the central pixel, it is not possible to fit for the anisotropy.  Therefore, we assume the anisotropy remain constant, that it is the same in the center as the best-fit value from the full FOV \cite{atlas3d15}\footnote{\texttt{Available from http://purl.org/atlas3d}}. This assumption is not necessarily true as anisotropy can vary with radius \citep[e.g.][]{gebhardt11,yildirim15}. 
\par To test if the variation in anisotropies could possibly explain the enhancement in $(M/L)_{\rm cen}$ relative to $(M/L)_{\rm dyn}$, we calculate the anisotropy that is required to eliminate this enhancement. Specifically, we create JAM models over a range of anisotropies until and find where the best-fit $(M/L)_{\rm cen}$ matches the $(M/L)_{\rm dyn}$, thus undoing the enhancement in the central $M/L$. We then compare the anisotropies required to eliminate this enhancement in each galaxy against the distribution of anisotropies inferred at larger radii.
\par
\begin{figure}[ht]

\epsscale{1.25}
\plotone {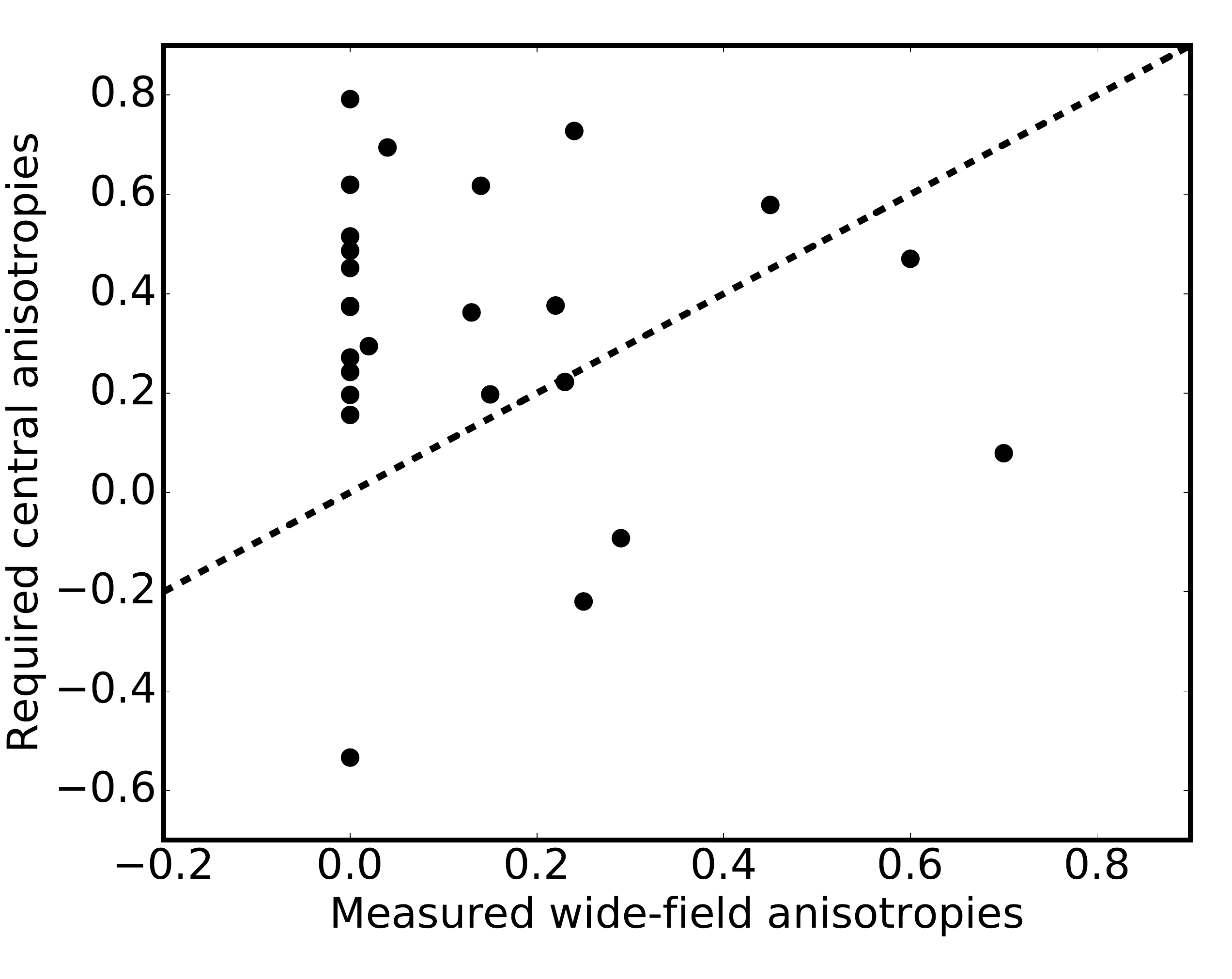}
\caption {Comparison of the anisotropies derived from model fits within 1$R_e$ \citep[x-axis][]{atlas3d15}, with the required anisotropies to eliminate the enhancement in the $(M/L)_{\rm cen}$ (y-axis). The distribution of anisotropies at larger radii have a median of just 0.09, while much higher values would be required to explain the enhancement in central $M/L$.}
\label{fig:13}

\end{figure}

 Figure~\ref{fig:13} shows the comparison between these two anisotropy values. The median of the anisotropies at large radii was found to be 0.09 for our sample of galaxies. We also calculated the median of anisotropies for the whole sample of 260 ATLAS$^{3D}$ galaxies which was found to be 0.05. Significantly higher radial anisotropy at the centers of these galaxies (median$\sim$0.6) would be required to explain the enhanced $(M/L)_{\rm cen}$s.  We think this large an enhancement in anisotropy at the centers of a majority of our galaxies is unlikely.

\section{Summary and Discussion}

In this work, we have examined the central dynamical mass-to-light ratios of a sample of 27 low-mass ($<$3$\times$10$^{10}$~M$_\odot$) early-type galaxies. For each galaxy, we construct a high-resolution luminosity/mass model from HST data. We combine this with the central velocity dispersion data from ATLAS$^{3D}$ to infer the central $(M/L)_{\rm cen}$ using JAM modeling. We exclude galaxies with dust features and strong color gradients that would affect the reliability of our modeling, which assumes mass traces light. We then compare our central $(M/L)_{\rm cen}$ values to expectations from stellar population models \{$(M/L)_{\rm pop}$\} and dynamical measurements at larger radii made by the ATLAS$^{3D}$ team \citep[$(M/L)_{\rm dyn},$][]{atlas3d20}. 

The results of our study are:
\begin{itemize}
\item The $(M/L)_{\rm cen}$ is elevated for $\sim$80\% of galaxies in the sample as compared to both $(M/L)_{\rm dyn}$ and $(M/L)_{\rm pop}$. This enhancement is statistically significant (3.3$\sigma$) over the full sample of galaxies, although many individual galaxies have only small enhancements. 
\item We examine systematic and random errors in our measurements and find that they cannot explain the enhancement we see. The elevation in the $(M/L)_{\rm cen}$ can be best described by a combination of central black holes or radial IMF variations within these galaxies. 
\item Two galaxies, NGC~4458 and NGC~4660, have $\gtrsim$3$\sigma$ enhancements in $(M/L)_{\rm cen}$ relative to both $(M/L)_{\rm dyn}$ and $(M/L)_{\rm pop}$. We suggest that these are due to BHs with a mass of $\sim$2$\times$10$^8$~M$_\odot$ in NGC~4458 and $\sim$8$\times$10$^8$~M$_\odot$ in NGC~4660.
\item Based on the comparison of $(M/L)_{\rm cen}$ to $(M/L)_{\rm dyn}$ and $(M/L)_{\rm pop}$, we estimate the BH masses (and upper limits) for each galaxy; in general these are upper limits above the expected values for the $M_\bullet-\sigma$ relationship.
\item We use the enhancements in $(M/L)_{\rm cen}$ relative to $(M/L)_{\rm dyn}$ to estimate a minimum and maximum scatter on the $M_\bullet-\sigma$ relationship of 0.4 and 1.2 dex at 95\% confidence. The upper limit is a firm one, as models with larger scatter would produce even more significant enhancements in $(M/L)_{\rm cen}$ than we observe.
\item The distribution of $(M/L)_{\rm cen}$ enhancements is not well explained just by black holes, therefore we consider the possibility that they are due to radial IMF variations; small variations in the IMF can account for the typical $\sim$14\% enhancement seen in the $(M/L)_{\rm cen}$ values relative to the $(M/L)_{\rm dyn}$ calculated at larger radii.
\end{itemize}

We now return to the discussion in Section~\ref{sec:ucds} on whether low-mass ETGs can act as the progenitors for UCDs. Putting the BHs observed at the centers of UCDs \citep{seth14,ahn17} into typical ETGs reduces the enhancement in velocity dispersion and inferred $M/L$.  Nonetheless, in our simulation adding a galaxy around M60-UCD1, we find an expected enhancement of $\sim$12\% in the $M/L$ due to the presence of a BH. This enhancement is very comparable to the median enhancement of $\sim$14\% that we see in our sample of low-mass ETGs. This supports the fact that UCDs can in fact be considered as nuclear remnants of ETGs.

\par Very few BHs are known in ETGs with masses~$\lesssim$~$10^{10}$~M$_\odot$ and $\sigma \lesssim 100$~km~s$^{-1}$. We have shown that deriving the central $M/L$ for nearby low-mass ETGs galaxies from ground-based spectroscopy can be a good probe for finding possible BHs. While in individual galaxies, we are sensitive only to BHs significantly more massive than expected from the $M_\bullet-\sigma$ relation, the overall distribution of central $M/L$ is sensitive to the scatter in this relation. 
We plan to follow up this work with adaptive optics observations of some of the most promising massive BH candidates, as well as examination of these galaxies at X-ray and radio wavelengths \citep[e.g.][]{gallo10}. We also find some evidence for radial IMF variations; more conclusive evidence could be gained from spectroscopic observations of IMF sensitive features across these galaxies.

{\em Acknowledgments:} RP and ACS acknowledge support for this work from HST AR-14313. ACS is also supported by NSF AST-1350389. JS acknowledges support from NSF grant AST-1514763 and the Packard Foundation. We thank Frank van den Bosch for his help with data on NGC~4342. Based on observations made with the NASA/ESA Hubble Space Telescope, and obtained from the Hubble Legacy Archive, which is a collaboration between the Space Telescope Science Institute (STScI/NASA), the Space Telescope European Coordinating Facility (ST-ECF/ESA) and the Canadian Astronomy Data Centre (CADC/NRC/CSA).

\bibliography{enhanced_ml.bib}
 
\newpage

\appendix
\section{List of parameters for the sample of galaxies}

\begin{table}[ht]
\centering
\caption{}
\label{table:mlr}
\begin{threeparttable}
\bgroup
\def\arraystretch{1.5}
\setlength\tabcolsep{2pt}
\begin{tabular}{ccccccccc}
\hline\hline
Galaxy     & $(M/L)_{\rm cen}$(Low - High)     & $(M/L)_{\rm dyn}$ & Quality         & $(M/L)_{\rm pop}$   &$\sigma_{\circ}$ & Camera    & Filter (for models)  & Filters (for colors) \\
			&($M_\odot/L_\odot$) 				&	($M_\odot/L_\odot$)		   & for $(M/L)_{\rm dyn}$   & ($M_\odot/L_\odot$) &(km s$^{-1}$)                                      \\
(1)         & (2)						& (3)			& (4)			& (5)		&(6)	& (7)		 & (8)    & (9)\\
\hline
NGC 2549    & 4.80 (4.51 - 5.11)         & 4.66        & 3               & 3.42   & 149.80 $\pm$ 2.79    & WFPC2/PC  & F702W  & --                  \\
NGC 2699    & 4.48 (4.06 - 4.91)         & 3.46        & 1               & 3.84   & 151.67 $\pm$ 5.30    & WFPC2/PC  & F702W  & --                  \\
NGC 2778    & 7.42 (6.98 - 7.87)         & 6.95        & 2               & 4.23   & 183.63 $\pm$ 3.22    & WFPC2/PC  & F814W  & F814W, F555W         \\
NGC 3412    & 1.87 (1.71 - 2.04)         & 1.64        & 1               & 2.38   & 103.94 $\pm$ 3.34    & WFPC2/PC  & F606W  & --                  \\
NGC 3522    & 3.07 (2.54 - 3.64)         & 3.86        & 0               & 2.73   & 101.76 $\pm$ 7.89    & WFC3/UVIS & F814W  & F814W, F475W         \\
NGC 3796    & 2.15 (1.87 - 2.44)         & 1.78        & 0               & 1.43   & 94.35 $\pm$ 5.07     & WFPC2/WF3 & F814W  & F814W, F606W         \\
NGC 4262    & 5.60 (5.30 - 5.90)         & 5.63        & 2               & 4.17   & 207.69 $\pm$ 2.96    & ACS/WFC   & F475W  & --                  \\
NGC 4339    & 6.09 (5.46 - 6.74)         & 4.83        & 0               & 3.87   & 138.57 $\pm$ 5.57    & WFPC2/PC  & F606W  & --                  \\
NGC 4342    & 11.14 (10.2 - 12.1)        & 10.43       & 3               & 4.23   & 317.10 $\pm$ 10.60$\footnote{This central velocity dispersion value is taken from \citet{vandenbosch98}}$   & WFPC2/PC  & F814W  & F814W, F555W         \\
NGC 4377    & 3.63 (3.37 - 3.89)         & 1.97        & 0               & 3.75   & 152.34 $\pm$ 3.48    & WFPC2/PC  & F606W  & F850LP, F475W        \\
NGC 4379    & 4.90 (4.49 - 5.33)         & 3.99        & 1               & 3.67   & 130.97 $\pm$ 3.97    & WFPC2/PC  & F814W  & F814W, F555W         \\
NGC 4387    & 2.84 (2.50 - 3.19)         & 3.75        & 1               & 3.68   & 92.90 $\pm$ 4.51     & ACS/WFC   & F475W  & F850LP, F475W        \\
NGC 4434    & 3.86 (3.65 - 4.08)         & 2.61        & 1               & 4.00   & 140.69 $\pm$ 2.09    & ACS/WFC   & F475W  & F850LP, F475W        \\
NGC 4458    & 4.30 (3.92 - 4.68)         & 2.88        & 1               & 3.57   & 120.69 $\pm$ 3.80    & WFPC2/PC  & F814W  & F814W, F555W         \\
NGC 4474    & 2.42 (1.97 - 2.90)         & 3.03        & 2               & 3.07   & 88.08 $\pm$ 7.37     & WFPC2/PC  & F702W  & F850LP, F475W        \\
NGC 4483    & 3.22 (2.66 - 3.82)         & 3.01        & 1               & 3.45   & 88.27 $\pm$ 6.86     & ACS/WFC   & F475W  & F850LP, F475W        \\
NGC 4489    & 1.34 (0.99 - 1.72)         & 0.98        & 0               & 1.74   & 55.49 $\pm$ 6.85     & ACS/WFC   & F475W  & F850LP, F475W        \\
NGC 4528    & 2.85 (2.54 - 3.16)         & 3.80        & 2               & 2.88   & 112.12 $\pm$ 4.68    & ACS/WFC   & F475W  & F850LP, F475W        \\
NGC 4551    & 3.56 (2.87 - 4.31)         & 4.98        & 2               & 3.89   & 97.33 $\pm$ 8.59     & ACS/WFC   & F475W  & F850LP, F475W        \\
NGC 4612    & 1.63 (1.39 - 1.87)         & 1.50        & 0               & 1.52   & 82.91 $\pm$ 5.01     & WFPC2/PC  & F606W  & F850LP, F475W        \\
NGC 4623    & 3.02 (2.28 - 3.85)         & 2.10        & 1               & 2.86   & 61.15 $\pm$ 7.15     & ACS/WFC   & F475W  & F850LP, F475W        \\
NGC 4638    & 3.06 (2.80 - 3.33)         & 2.71        & 3               & 3.65   & 132.29 $\pm$ 4.06    & ACS/WFC   & F475W  & F850LP, F475W        \\
NGC 4660    & 5.96 (5.64 - 6.27)         & 4.00        & 2               & 4.19   & 232.18 $\pm$ 3.24    & WFPC2/PC  & F814W  & F814W, F555W         \\
NGC 4733    & 2.79 (1.81 - 3.94)         & 2.18        & 0               & 1.69   & 49.85 $\pm$ 8.86     & WFPC2/PC  & F814W  & F814W, F555W        \\
NGC 7457    & 2.03 (1.67 - 2.42)         & 1.80        & 1               & 1.34   & 65.11 $\pm$ 5.16     & WFPC2/PC  & F814W  & F814W, F555W        \\
PGC 028887  & 5.05 (4.57 - 5.53)         & 4.50        & 0               & 3.72   & 135.84 $\pm$ 4.78    & WFC3/UVIS & F814W  & F814W, F475W        \\
PGC 050395  & 2.46 (1.89 - 3.09)         & 1.04        & 0               & 1.91   & 77.42 $\pm$ 8.40     & WFC3/UVIS & F814W  & F814W, F475W        \\
\hline
\end{tabular}

\begin{tablenotes}
$Note$: Column (1): Galaxy's name from LEDA \citep{paturel03}, which is used as a standard here. Column (2): Central M/L derived using JAM models. All the values are in SDSS $r$-band. The Low and High values include the errors from velocity dispersions and 2.5\% systematic error. Column (3): Dynamical $M/L$ derived by dynamically fitting the outer regions of the galaxy ($>2''$, within $1Re$). We assume an error of 6\% on all of these $M/L$ as given in \citep{atlas3d20}. Column (4): Qual=0 indicates a low-quality JAM fit that is due to barred galaxies, dust or uncertain deprojection due to a low inclination of the galaxy. Qual=3 indicates a good JAM fit to the data. Column (5): Stellar population $M/L$ derived by fitting the ATLAS$^{3D}$ central spectrum of a galaxy assuming a Kroupa IMF. We assume an error of 6\% on all of these $M/L$ as given in \citep{atlas3d20}. Column (6): Central velocity dispersions derived from ATLAS$^{3D}$ kinematic data. Column (7) \& (8): Images of galaxies from the HST camera and their corresponding filters used in deriving the mass models. Column (9): Filters that were used in determining the color maps of the galaxies. 
\end{tablenotes}
\egroup
\end{threeparttable}

\end{table}

\begin{table}[ht]
\centering
\caption{}
\label{table:bhmass}
\begin{threeparttable}
\bgroup
\def\arraystretch{1.5}
\setlength\tabcolsep{2pt}

\begin{tabular}{>{\rowmac}c>{\rowmac}c>{\rowmac}c>{\rowmac}c>{\rowmac}c>{\rowmac}c>{\rowmac}c>{\rowmac}c>{\rowmac}c<{\clearrow}}

\hline\hline
Galaxy & $\sigma_e$   & log$M_{\star}$ & M$_{\bullet,\rm dyn}$(Low-High) & M$_{\bullet,\rm pop}$(Low-High)& (M-${\sigma}$)$_{\bullet}$ &$r_{SOI}$ & Pixel size & Pixel size\\
       & (km s$^{-1}$) & ($M_{\odot}$)  & ($M_{\odot}$)                   & (M$_{\odot}$)   				  & (M$_{\odot}$)	           &(pc) & ATLAS$^{3D}$(pc) &HST (pc)\\
   (1) & (2)		  & (3)			   & (4)						   	 & (5)							  & (6)                        &(7) &(8) &(9)\\
\hline
NGC 2549 & 2.152 & 10.435 & 3.37 (0.00 - 16.5)$\times$10$^7$  & 3.14 (2.00 - 4.28)$\times$10$^8$ & 3.31$\times$10$^7$ & 7.26   & 56.05  & 2.98\\
NGC 2699 & 2.104 & 10.387 & 3.79 (1.47 - 6.15)$\times$10$^8$  & 2.36 (0.00 - 4.81)$\times$10$^8$ & 1.85$\times$10$^7$ & 101.68 & 119.40 & 6.35\\
NGC 2778 & 2.121 & 10.497 & 1.77 (0.00 - 5.00)$\times$10$^8$  & 1.19 (0.93 - 1.45)$\times$10$^9$ & 2.27$\times$10$^7$ & 43.97  & 101.63 & 5.41\\
NGC 3412 & 1.967 & 10.162 & 6.29 (0.00 - 13.4)$\times$10$^7$  & 0.00 (0.00 - 3.82)$\times$10$^7$ & 3.54$\times$10$^6$ & 31.72  & 50.13  & 2.67\\
NGC 3522 & 1.993 & 10.305 & 0.00 (0.00 - 5.13)$\times$10$^7$  & 7.57 (0.00 - 23.9)$\times$10$^7$ & 4.84$\times$10$^6$ & 22.94  & 116.21 & 4.95\\
NGC 3796 & 1.916 & 9.987  & 5.93 (0.00 - 12.3)$\times$10$^7$  & 1.14 (0.56 - 1.74)$\times$10$^8$ & 1.91$\times$10$^6$ & 37.84  & 103.45 & 11.01\\
NGC 4262 & 2.207 & 10.479 & 0.00 (0.00 - 2.03)$\times$10$^8$  & 8.59 (5.30 - 11.9)$\times$10$^8$ & 6.43$\times$10$^7$ & 33.89  & 70.18  & 3.73\\
NGC 4339 & 1.98  & 10.433 & 2.39 (0.65 - 4.17)$\times$10$^8$  & 4.18 (2.55 - 5.86)$\times$10$^8$ & 4.14$\times$10$^6$ & 113.41 & 72.92  & 3.88\\
NGC 4342 & 2.384 & 10.521 & 1.33 (0.00 - 4.28)$\times$10$^8$  & 1.29 (1.07 - 1.51)$\times$10$^9$ & 5.45$\times$10$^8$ & 9.84   & 75.19  & 4.00\\
NGC 4377 & 2.092 & 10.246 & 5.04 (3.87 - 6.13)$\times$10$^8$  & 0.00 (0.00 - 6.77)$\times$10$^7$ & 1.60$\times$10$^7$ & 142.85 & 81.12  & 4.31\\
NGC 4379 & 1.994 & 10.269 & 2.06 (0.59 - 3.55)$\times$10$^8$  & 2.78 (1.35 - 4.22)$\times$10$^8$ & 4.90$\times$10$^6$ & 91.78  & 72.00  & 3.83\\
NGC 4387 & 1.998 & 10.182 & 0.00 (0.00 - 3.89)$\times$10$^7$  & 0.00 (0.00 - 3.82)$\times$10$^7$ & 5.15$\times$10$^6$ & 16.99  & 81.57  & 4.34\\
NGC 4434 & 1.999 & 10.21  & 4.86 (3.44 - 6.29)$\times$10$^8$  & 0.00 (0.00 - 9.27)$\times$10$^7$ & 5.21$\times$10$^6$ & 211.39 & 102.08 & 5.43\\
\setrow{\bfseries} NGC 4458 & 1.947 & 10.032 & 2.01 (1.24 - 2.80)$\times$10$^8$& 1.03 (0.20 - 1.87)$\times$10$^8$ & 2.78$\times$10$^6$ & 111.30 & 74.74 & 3.98 \\
NGC 4474 & 1.93  & 10.187 & 0.00 (0.00 - 4.82)$\times$10$^7$ & 0.00 (0.00 - 4.88)$\times$10$^7$ & 2.26$\times$10$^6$ & 28.82 & 71.09 & 3.78 \\
NGC 4483 & 1.941 & 10.138 & 4.65 (0.00 - 21.5)$\times$10$^7$ & 0.00 (0.00 - 4.47)$\times$10$^7$ & 2.59$\times$10$^6$ & 26.42 & 76.11 & 4.05 \\
NGC 4489 & 1.832 & 9.866  & 5.07 (0.00 - 11.3)$\times$10$^7$ & 0.00 (0.00 - 1.46)$\times$10$^7$ & 6.93$\times$10$^5$ & 47.58 & 70.18 & 3.73 \\
NGC 4528 & 2.007 & 10.123 & 0.00 (0.00 - 5.28)$\times$10$^7$ & 0.00 (0.00 - 4.01)$\times$10$^7$ & 5.74$\times$10$^6$ & 22.13 & 72.00 & 3.83 \\
NGC 4551 & 1.971 & 10.262 & 0.00 (0.00 - 2.64)$\times$10$^7$ & 0.00 (0.00 - 2.07)$\times$10$^7$ & 3.71$\times$10$^6$ & 13.06 & 73.37 & 3.90 \\
NGC 4612 & 1.935 & 10.265 & 3.69 (0.00 - 13.0)$\times$10$^7$ & 3.08 (0.00 - 12.4)$\times$10$^7$ & 2.40$\times$10$^6$ & 21.54 & 75.65 & 4.02 \\
NGC 4623 & 1.884 & 10.17  & 7.91 (0.50 - 16.0)$\times$10$^7$ & 1.36 (0.00 - 9.86)$\times$10$^7$ & 1.30$\times$10$^6$ & 58.43 & 79.30 & 4.22 \\
NGC 4638 & 2.134 & 10.408 & 1.60 (0.00 - 3.54)$\times$10$^8$ & 0.00 (0.00 - 9.87)$\times$10$^7$ & 2.66$\times$10$^7$ & 37.40 & 79.75 & 4.24 \\
\setrow{\bfseries} NGC 4660 & 2.263 & 10.499 & 7.78 (5.57 - 9.98)$\times$10$^8$ & 6.98 (4.72 - 9.22)$\times$10$^8$ & 1.26$\times$10$^8$ & 100.32 & 68.36 & 3.64 \\
NGC 4733 & 1.717 & 9.877  & 4.64 (0.00 - 14.3)$\times$10$^7$ & 8.32 (0.20 - 17.8)$\times$10$^7$ & 1.73$\times$10$^5$  & 73.90  & 66.08  & 3.51\\
NGC 7457 & 1.873 & 10.218 & 1.99 (0.00 - 6.11)$\times$10$^7$ & 5.77 (2.08 - 9.67)$\times$10$^7$ & 1.14$\times$10$^6$  & 15.44  & 58.79  & 3.13\\
PGC 028887 & 2.109 & 10.534 & 1.75 (0.00 - 4.16)$\times$10$^8$ & 4.21 (1.99 - 6.49)$\times$10$^8$ & 1.97$\times$10$^7$  & 45.84  & 186.85 & 7.95\\
PGC 050395 & 1.909 & 10.145 & 4.00 (2.23 - 5.94)$\times$10$^8$ & 1.55 (0.00 - 3.63)$\times$10$^8$ & 1.76$\times$10$^6$  & 263.61 & 169.53 & 7.21\\

\hline 

\end{tabular}
\begin{tablenotes}
$Note$: Column (1): Galaxy's name from LEDA \citep{paturel03}, which is used as a standard here. Column (2): Effective stellar velocity dispersion adopted from \citep{atlas3d15}. They were calculated by co-adding all SAURON spectra within an effective ellipse of area $\pi R_e^2$, where $R_e$ is the effective radius. Column (3): Galaxy mass as taken from \citet{atlas3d15}. Column (4): Best-fit BH masses derived using $(M/L)_{\rm dyn}$ as the true stellar $M/L$ in our JAM models. Column (5): Best-fit BH masses derived using $(M/L)_{\rm pop}$ as the true stellar $M/L$ in our JAM models.  Column (6): Expected BH masses from the M$_\bullet-\sigma$ relationship from Equation~\ref{msigma} . Galaxies that are highlighted are highly significant detections. Column (7): Sphere of influence calculated from the derived BH masses (column 4) using $r_{SOI} = G.M_\bullet/\sigma_e^2$. Column (8): Pixel size of ATLAS$^{3D}$ kinematic data for our sample of galaxies that were derived using the distances from \citet{atlas3d1}.  Column (9): Pixel size of HST images for our sample of galaxies that were derived using the distances from \citet{atlas3d1}.
\end{tablenotes}
\egroup
\end{threeparttable}
\end{table}

\end{document}